\documentclass[A4,12pt]{article}

\usepackage{cite}

\usepackage{amsfonts}
\usepackage{pict2e}

\title{Massive higher spin fields\\
in the frame-like multispinor formalism}

\author{M.V. Khabarov${}^{ab}$\thanks{maksim.khabarov@ihep.ru},
Yu.M. Zinoviev${}^{ab}$\thanks{Yurii.Zinoviev@ihep.ru}
\\[0.5cm]
\it{\small ${}^a$Institute for High Energy Physics of National
Research Center "Kurchatov Institute"} \\
\it{\small Protvino, Moscow Region, 142281, Russia} \\
\it\small{ ${}^b$Moscow Institute of Physics and Technology (State
University),} \\
\it{\small Dolgoprudny, Moscow Region, 141701, Russia}}

\date{}

\begin{document}

\maketitle

\begin{abstract}
In this paper, a gauge invariant description of massive higher spin
bosonic and fermionic particles in frame-like Lagrangian and unfolded
formalism in (A)dS${}_4$ is built. A complete set of gauge invariant
object is also constructed and the Lagrangian is rewritten in terms of
these objects. The unitarity of the theories is studied alongside with
the partially massless limits. The calculations are carried out in the
multispinor formalism, which simplifies them and is particularly
convenient for the supersymmetry studies.
\end{abstract}

\thispagestyle{empty}
\newpage
\setcounter{page}{1}
\tableofcontents\pagebreak


\section{Introduction}

There are two well known formalisms for the description of the
massless higher spin fields: the metric one
\cite{Fro78,FF78,Fro79,FF79}, which can be considered as a higher spin
generalization of the metric formulation of gravity, and a so-called
frame-like one \cite{Vas80,LV88,Vas88} generalizing a frame
formulation for gravity. Both formalisms are drastically based on the
gauge invariance which guarantees the correct number of the physical
degrees of freedom and almost completely fixes the possible forms of
consistent interactions.

The metric formulation of the massive bosonic and fermionic fields was
proposed long ago in \cite{SH74,SH74a}. It does not possess any gauge
invariance; instead, it provides us with the set of the constraints
which follow from the Lagrangian equations. One of the possible routes
for the investigation of the consistent interactions for the massive
higher spin fields is to use their gauge invariant description, which
in the metric approach has been proposed in \cite{Zin01,Met06}. The
construction of the frame-like gauge invariant formulation for the
massive higher spin bosons and fermions was initiated in
\cite{Zin08b}, while the full fledged formulation for the bosonic case
in arbitrary space-time dimensions $d \ge 4$ was developed in
\cite{PV10}. Later on it was shown that such formalism can also be
used for the description of the infinite spin fields as well 
\cite{Met16,Met17,KhZ17}.

The tensor formulation used in \cite{PV10} is universal in a sense
that it works in any space-time dimensions $d \ge 4$, but it appears
technically quite involved. It becomes even more complicated in the
case of the massive fermions and this is, at least, one of the reasons
why such formalism has not been developed so far. In this work we
restrict ourselves with the four dimensional  space-time. This allows
us to use a multispinor formalism which greatly simplifies
calculations especially when one has to deal with the mixed symmetry
(spin-)tensors. So we managed not only reproduce the results of
\cite{PV10} (with a number of generalizations) but also developed an
analogous formulation for the massive fermions. Note, that such
formalism, where bosons and fermions appear on equal footing, is very
well suited for the investigation of the supersymmetric models.
Indeed, it has been already used in our recent investigations of
different $N=1$ supermultiplets in four dimensions
\cite{BKhSZ19,BKhSZ19a,BKhSZ19b}.

In this paper, we develop two different but tightly connected
formalisms - namely, the frame-like Lagrangian one and the unfolded
one. Let us illustrate them on the simple case of the massless 
spin-$s$ field propagating over the $(A)dS_4$ background. In the
frame-like multispinor formalism we use here (see Appendix A for
notations and conventions), the massless spin-$s$ boson is described
by the physical one-form $\Phi^{\alpha(s-1)\dot\alpha(s-1)}$ and
auxiliary one-forms $\Omega^{\alpha(s)\dot\alpha(s-2)} + h.c.$. The
free Lagrangian (which is a four-form in our formalism) looks as
follows:
\begin{eqnarray}
-i(-1)^{s}\mathcal{L} &=&
s \Omega^{\alpha(s-2)\dot\gamma\dot\alpha(s-1)} 
E_{\dot\gamma\dot\beta} 
\Omega_{\alpha(s-2)}{}^{\dot\beta}{}_{\dot\alpha(s-1)}
- (s-2)\Omega^{\alpha(s-3)\gamma\dot\alpha(s)}
E_{\gamma\beta}\Omega_{\alpha(s-3)}{}^\beta{}_{\dot\alpha(s)}
\nonumber
\\
&& + 2\Omega^{\alpha(s-2)\dot\gamma\dot\alpha(s-1)}e_{\beta\dot\gamma}
D \Phi_{\alpha(s-2)}{}^\beta{}_{\dot\alpha(s-1)} \nonumber 
\\
 && + 2 \lambda^2\Phi^{\alpha(s-1)\dot\alpha(s-1)}
E_\alpha{}^\beta\Phi_{\alpha(s-2)\beta\dot\alpha(s-1)} - h.c.
\end{eqnarray}
This Lagrangian is invariant under the following gauge
transformations:
\begin{eqnarray}
\delta \Omega^{\alpha(s-2)\dot\alpha(s)} &=& 
D\eta^{\alpha(s-2)\dot\alpha(s)}
+ (s-2)e^\alpha{}_{\dot\alpha} \zeta^{\alpha(s-3)\dot\alpha(s+1)}
+ s\lambda^2 e_\alpha{}^{\dot\alpha} \xi^{\alpha(s-1)\dot\alpha(s-1)},
\nonumber
\\
\delta \Phi^{\alpha(s-1)\dot\alpha(s-1)} &=&
D \xi^{\alpha(s-1)\dot\alpha(s-1)}
+ (s-1) e^\alpha{}_{\dot\alpha} \eta^{\alpha(s-2)\dot\alpha(s)}
+ (s-1) e_\alpha{}^{\dot\alpha} \eta^{\alpha(s)\dot\alpha(s-2)}.
\label{itrans}
\end{eqnarray} 
It is easy to construct a gauge invariant two-form (an analogue of the
torsion in gravity):
\begin{equation}
R^{\alpha(s-1)\dot\alpha(s-1)} =
D \Phi^{\alpha(s-1)\dot\alpha(s-1)}
+ (s-1) e^\alpha{}_{\dot\alpha} \Omega^{\alpha(s-2)\dot\alpha(s)}
+ (s-1) e_\alpha{}^{\dot\alpha} \Omega^{\alpha(s)\dot\alpha(s-2)}.
\end{equation} 
The straightforward attempt to generalize the curvature by
substituting the frame and the spin-connection with the physical and
the auxiliary fields respectively, however, fails: the result is not
invariant under the part of the transformation (\ref{itrans})
parametrized by $\zeta^{\alpha(s+1)\dot\alpha(s-3)}$, which has no
analogue in  the spin-2 case. To restore the full invariance, one has
to introduce a so-called extra field 
$\Sigma^{\alpha(s+1)\dot\alpha(s-3)}$ (with its complex
conjugate), which does not enter the free Lagrangian (although it is
required to build the gauge invariant interactions) and plays the
role of the gauge field for this extra gauge transformations:
\begin{eqnarray}
\delta \Sigma^{\alpha(s+1)\dot\alpha(s-3)} &=&
D \zeta^{\alpha(s+1)\dot\alpha(s-3)}
+ (s-3)e_\beta{}^{\dot\alpha} \zeta^{\alpha(s+1)\beta\dot\alpha(s-4)}
\nonumber
\\
 && + (s+1)\lambda^2 e^\alpha{}_{\dot\beta}
\eta^{\alpha(s)\dot\beta\dot\alpha(s-3)}.
\end{eqnarray}
With the use of this extra field, one builds the generalization of the
Riemann tensor as:
\begin{equation}
R^{\alpha(s)\dot\alpha(s-2)} =
D \Omega^{\alpha(s)\dot\alpha(s-2)}
+ s\lambda^2 e^\alpha{}_{\dot\beta}
\Phi^{\alpha(s-1)\dot\beta\dot\alpha(s-2)}
+ (s-2) e_\alpha{}^{\dot\alpha}
\Sigma^{\alpha(s+1)\dot\alpha(s-3)}.
\end{equation}
It is possible to construct a gauge invariant object which contains
the derivative of the extra field
$\Sigma^{\alpha(s+1)\dot\alpha(s-3)}$; however, the extra field
$\Sigma^{\alpha(s+2)\dot\alpha(s-4)}$ is needed for that. In the
end, one arrives at the complete set of extra fields:
\begin{eqnarray}
\delta \Sigma^{\alpha(s-1+m)\dot\alpha(s-1-m)} &=&
D \zeta^{\alpha(s-1+m)\dot\alpha(s-1-m)}
+ (s-1-m) e_\beta{}^{\dot\alpha}
\zeta^{\alpha(s-1+m)\beta\dot\alpha(s-2-m)} \nonumber
\\
 && + (s-1+m)\lambda^2 e^\alpha{}_{\dot\beta}
\zeta^{\alpha(s-2+m)\dot\beta\dot\alpha(s-1-m)}.
\end{eqnarray}
used to build the complete set of the gauge invariant curvatures:
\begin{eqnarray}
R^{\alpha(s-1+m)\dot\alpha(s-1-m)} &=&
D \Sigma^{\alpha(s-1+m)\dot\alpha(s-1-m)}
+ (s-1+m)\lambda^2 e^\alpha{}_{\dot\alpha}
\Sigma^{\alpha(s-2+m)\dot\alpha(s-m)} \nonumber
\\
 && + (s-1-m)e_\alpha{}^{\dot\alpha}
\Sigma^{\alpha(s+m)\dot\alpha(s-2-m)}.
\end{eqnarray}
The index $|m|\le s-1$; in case of $|m|=s-1$ the terms with zero
coefficients $(s-1\pm m)$ are omitted. This set of curvatures is
closed, i.e. for each field $\Sigma^{\alpha(s-1+m)\dot\alpha(s-1-m)}$
there is a unique curvature $R^{\alpha(s-1+m)\dot\alpha(s-1-m)}=
DW^{\alpha(s-1+m)\dot\alpha(s-1-m)}+\ldots$. Note the differential
relation
\begin{eqnarray}
D R^{\alpha(s-1+m)\dot\alpha(s-1-m)} &=&
- (s-1+m)\lambda^2 e^\alpha{}_{\dot\alpha}
R^{\alpha(s-2+m)\dot\alpha(s-m)} \nonumber
\\
 && - (s-1-m) e_\alpha{}^{\dot\alpha}
R^{\alpha(s+m)\dot\alpha(s-2-m)}.
\label{iRel}
\end{eqnarray}
As in the case of gravity, the Lagrangian can be rewritten in the
manifestly gauge invariant form:
\begin{eqnarray}
i(-1)^{s+1}\mathcal{L} &=&
\sum_{m=1}^{s-1} \frac{(s-2)!(s-1)!}{(s-1-m)!(s+m-1)! \lambda^{2m}}
R^{\alpha(s-1+m)\dot\alpha(s-1-m)} R_{\alpha(s-1+m)\dot\alpha(s-1-m)}
\nonumber
\\
 && -h.c.
\label{iLag}
\end{eqnarray}
The coefficients in the (\ref{iLag}) are determined by the extra field
decoupling condition
\begin{equation}
\frac{\delta\mathcal{L}}{\delta
\Sigma^{\alpha(s-1+m)\dot\alpha(s-1-m)}} = 0
\end{equation}
up to the normalization factor.

We now turn on to describe of the unfolded formalism. It describes
the particle via an infinite chain of first order equations closed
on-shell. No Lagrangian is known that would imply the whole chain of
the equations. It is remarkable that the unfolded formalism is the
only one in which a complete non-linear theory has been constructed
\cite{Vas90,Vas91a,Vas92}. In \cite{PV10} the unfolded formulation was
constructed for the massive bosons; one of the aims of our paper is to
build the unfolded description for the massive fermions.

 We derive the unfolded equations now. We start with an anlogue of the
zero torsion condition: 
\begin{equation}
T^{\alpha(s-1)\dot\alpha(s-1)} = D
\Phi^{\alpha(s-1)\dot\alpha(s-1)} + (s-1)e_\alpha{}^{\dot\alpha}
\Omega^{\alpha(s)\dot\alpha(s-2)} + h.c. = 0,
\end{equation}
which holds on-shell and allows us to express 
$\Omega^{\alpha(s)\dot\alpha(s-2)}$ in terms of
$D\Phi^{\alpha(s-1)\dot\alpha(s-1)}$ up to the gauge transformations.
Its derivative can be expressed via $R^{\alpha(s)\dot\alpha(s-2)}$
(with its complex conjugate) using the identity (\ref{iRel}):
\begin{equation}
DT^{\alpha(s-1)\dot\alpha(s-1)} = - (s-1)e_\alpha{}^{\dot\alpha}
R^{\alpha(s)\dot\alpha(s-2)} + h.c. = 0.
\end{equation}
One can see that the condition $e_\alpha{}^{\dot\alpha}
R^{\alpha(s)\dot\alpha(s-2)} + h.c. =0$ ensures there is one-to-one
correspondence between the components of
$R^{\alpha(s)\dot\alpha(s-2)}$  and
$\Sigma^{\alpha(s+1)\dot\alpha(s-3)}$ up to the gauge invariance.
So we can set
\begin{equation}
0 = R^{\alpha(s){\dot{\alpha}}(s-2)} = 
s\lambda^2 e^\alpha{}_{\dot\alpha} \Phi^{\alpha(s-1)\dot\alpha(s-1)}
+ D \Omega^{\alpha(s)\dot\alpha(s-2)}
+ (s-2) e_\alpha{}^{\dot\alpha} \Sigma^{\alpha(s+1)\dot\alpha(s-3)},
\end{equation}
and solve it for the $\Sigma^{\alpha(s+1)\dot\alpha(s-3)}$. By
repeating the steps above, we arrive at the number of zero curvature
conditions: 
\begin{equation}
R^{\alpha(s+m-1)\dot\alpha(s-m-1)} = 0, \qquad |m| \le s-2
\end{equation}
The case of the curvature $R^{\alpha(2s-2)}$ is quite different. In
the previous steps, we always had an extra field which could be chosen
to set the curvature to zero. There is no extra field left when we
obtain $e_\alpha{}^{\dot\alpha} R^{\alpha(2s-2)}=0$. Thus to write the
most general consistent equation we have to introduce a first gauge
invariant zero-form:
\begin{equation}
R^{\alpha(2s-2)} = E_{\alpha(2)} W^{\alpha(2s)}
\end{equation}
It is the
closest analogue of the Weyl tensor in the gravity; it parametrizes
all the components which do not vanish on-shell. The derivative of the
curvature $R^{\alpha(2s-2)}$ can be expressed via other curvatures and
thus vanish. This gives the condition
\begin{equation}
E_{\alpha(2)} D W^{\alpha(2s)} = 0
\end{equation}  
for the zero-form $W^{\alpha(2s)}$. Similarly to the previous steps,
this means that its derivative can be uniquely expressed via the
components of another field. In this case, the field is
$W^{\alpha(2s+1)\dot\alpha}$:
\begin{equation}
0 = DW^{\alpha(2s)} + e^\alpha{}_{\dot\alpha} 
W^{\alpha(2s+1)\dot\alpha}.
\end{equation}  
In turn, the equation for the $W^{\alpha(2s+1)\dot\alpha}$ requires
introduction of $W^{\alpha(2s+2)\dot\alpha(2)}$ and so on. We obtain
an infinite chain of zero-forms $W^{\alpha(2s+m)\dot\alpha(m)} +
h.c.$, $m \ge 0$ (see Figure 1):
\begin{equation}
0 = DW^{\alpha(2s+m)\dot\alpha(m)} + e_{\alpha\dot\alpha}
W^{\alpha(2s+m+1)\dot\alpha(m+1)} + 
(2s+m)m\lambda^2e^{\alpha\dot\alpha}
W^{\alpha(2s+m-1)\dot\alpha(m-1)}
\end{equation}
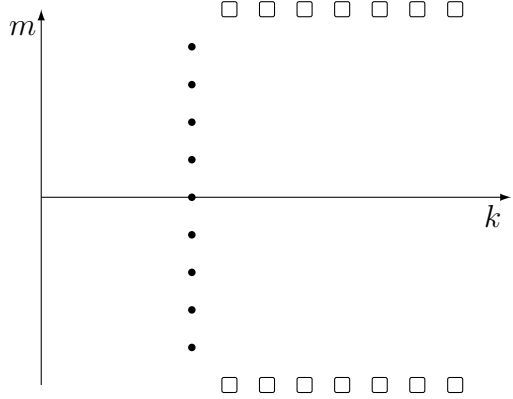
\begin{figure}[htb]
\begin{center}
\begin{picture}(140,120)

\put(10,10){\vector(0,1){100}}
\put(0,100){\makebox(10,10){$m$}}
\put(10,60){\vector(1,0){125}}
\put(125,50){\makebox(10,10){$k$}}

\multiput(50,20)(0,10){9}{\circle*{2}}

\multiput(60,110)(10,0){7}{\oval[0.5](4,4)}
\multiput(60,10)(10,0){7}{\oval[0.5](4,4)}

\end{picture}
\end{center}
\caption{Spectrum of one-forms (dots) and gauge invariant zero-forms
(squares) for the massless boson with $s=5$}
\end{figure}
One can see that each equality in the unfolded equations chain gives
the parametrization of the derivatives of the previous field not
vanishing on-shell by the next field (up to the $\lambda^2$ terms
induced by the space-time curvature). Hence, the $i$-th pair of the
one-forms (taking physical field as zeroth) represents all the $i$-th
derivatives of the physical field which do not vanish on-shell. The
same holds for the zero-forms - the $i$-th pair of zero-forms
represents the $(s+i)$-th derivative. This is an explanation why the
extra fields do not enter the free Lagrangian - it has second order in
the derivatives, if one expresses all the fields via the physical one.
Moreover, the fermionic Lagrangian has the first order and hence
contains the physical field only.

In the next sections, we build the frame-like and unfolded formulation
for massive higher spin particles. Our main interest here is the
general massive case, however we also investigate all possible
partially massless and/or infinite spin limits. The Sections 2
and 3 are devoted to bosons and fermions respectively. Each section is
divided into five parts. In the first part, we construct the gauge
invariant Lagrangian for the particle and study its unitarity. In the
second part, we build the complete set of the gauge invariant
curvatures, introducing all necessary extra fields. In the third
part, we use the set of the curvatures to express the Lagrangian in
the explicitly gauge invariant form. In the fourth part, we build the
unfolded equations. And in the final fifth part, we discuss the
applications of our work to the formalism developed in \cite{SV06}. In
Appendix A the notations, conventions and the facts about the
multispinor formalism are presented. In Appendix B we list the general
facts about the gauge invariant curvatures for the massive particles.

\section{Bosonic case}

\subsection{The Lagrangian}

To construct the gauge invariant description for the massive spin-$s$
boson, one has to introduce a complete set of components for the
massless fields with spins from zero to $s$ \cite{Zin01,Zin08b}. As
it was already mentioned in the introduction, one needs two fields to
describe the massless boson of spin $s>1$ --- namely, the physical
one-form $\Phi^{\alpha(k)\dot\alpha(k)}$ and the
auxiliary one-form $\Omega^{\alpha(k+1)\dot\alpha(k-1)}$ with its
complex conjugate. The cases with spins $s=1$ and $s=0$ are special:
one needs one-form $A$ and zero-form $B^{\alpha(2)} + h.c.$ for spin-1
and zero-forms $\phi$, $\pi^{\alpha\dot\alpha}$ for spin-0. The
complete Lagrangian is built as a sum of kinetic terms for all fields
with all possible cross and mass terms added. The most general ansatz
(up to the normalization choice) is: 
\begin{equation}
\label{bLag}
\mathcal{L} = \mathcal{L}_0 + \mathcal{L}_1 + \mathcal{L}_2
\end{equation}
\begin{eqnarray}
-i\mathcal{L}_0 &=& \sum_{k=1}^{s-1}
(-1)^{k+1}\big[[(k+1)\Omega^{\alpha(k-1)\dot\gamma\dot\alpha(k)}E_{\dot\gamma\dot\beta} 
\Omega_{\alpha(k-1)}{}^{\dot\beta}{}_{\dot{\alpha}(k)} \nonumber
\\
 && \qquad \qquad -
(k-1)\Omega^{\alpha(k-2)\gamma\dot\alpha(k+1)}E_{\gamma\beta}
\Omega_{\alpha(k-2)}{}^\beta{}_{\dot\alpha(k+1)} ] \nonumber
\\
 && \qquad \qquad +
2\Omega^{\alpha(k-1)\dot\gamma\dot\alpha(k)}e_{\beta\dot\gamma}
D\Phi_{\alpha(k-1)}{}^\beta{}_{\dot\alpha(k)} \big] - h.c. \nonumber
\\
 && - 4\mu_1^2 E B^{\alpha(2)} B_{\alpha(2)}
 - 2\mu_1E_{\alpha(2)} B^{\alpha(2)} DA - h.c. \nonumber
\\
 && +12\mu_1^2\beta_2 E\pi^{\alpha\dot\alpha}\pi_{\alpha\dot\alpha}
- 24\mu_1 \beta_2 E_{\alpha\dot\alpha} \pi^{\alpha\dot\alpha} D
\phi, 
\\
-i\mathcal{L}_1 &=& \sum_{k=2}^{s-1} (-1)^{k+1}\big[
- \frac{2(k+1)\mu_k}{(k-1)}
\Omega^{\alpha(k+1)\dot\alpha(k-1)}E_{\alpha(2)}
\Phi_{\alpha(k-1)\dot\alpha(k-1)} \nonumber
\\
 && \qquad \qquad +
2\mu_k\Omega_{\alpha(k-2)\dot\alpha(k)}E_{\alpha(2)}
\Phi^{\alpha(k)\dot\alpha(k)} - h.c.\big] \nonumber
\\
 && - 2\mu_1\Omega^{\alpha(2)} E_{\alpha(2)} A
- 4\mu_1^2E_{\alpha\dot\alpha} B^\alpha{}_\beta\Phi^{\beta\dot\alpha} 
+h.c. + 24\beta_2\mu_1E_{\alpha\dot\alpha} \pi^{\alpha\dot\alpha} A,
\\
-i\mathcal{L}_2 &=& \sum_{k=1}^{s-1} (-1)^{k+1} \big[
2\beta_{k+1}(k+1)\Phi^{\alpha(k)\dot\alpha(k)}
E_\alpha{}^\beta\Phi_{\alpha(k-1)\beta\dot\alpha(k)} - h.c. \big]
\nonumber
\\
 && -24\mu_1\beta_2E_{\alpha\dot\alpha} \Phi^{\alpha\dot\alpha}
\phi+24 \mu_1^2\beta_2 E \phi^2.
\end{eqnarray}
Here all the terms are arranged into three sums
$\mathcal{L}_0$, $\mathcal{L}_1$, $\mathcal{L}_2$ by the
dimensionality of the coefficients. To simplify the calculations, a
non-canonical normalization of the fields $B^{\alpha(2)}$, 
$\pi^{\alpha\dot\alpha}$ and $\phi$ is chosen. The
same Lagrangian can also be used to describe the infinite spin
particle by taking $s \to \infty$ \cite{Met16,KhZ17}. 

In order to have the right amount of the physical degrees of freedom,
the Lagrangian has to possess all the symmetries of the initial
massless Lagrangians. In this, the gauge transformations also have to
be modified with cross and mass-like terms. The ansatz for the 
transformations consistent with the Lagrangian has the form:
\begin{eqnarray}
\delta\Omega^{\alpha(k+1)\dot\alpha(k-1)} &=&
D\eta^{\alpha(k+1)\dot\alpha(k-1)}
+ (k-1) e_\alpha{}^{\dot\alpha}
\zeta^{\alpha(k+2)\dot\alpha(k-2)}
+ (k+1)\beta_{k+1} e^\alpha{}_{\dot\alpha}
\xi^{\alpha(k)\dot\alpha(k)} \nonumber
\\
 && + \frac{(k+1)}{(k+2)}\mu_{k} e^{\alpha\dot\alpha}
\eta^{\alpha(k)\dot\alpha(k-2)}
+ \frac{(k+2)}{k}\mu_{k+1} e_{\alpha\dot\alpha}
\eta^{\alpha(k+2)\dot\alpha(k)}, \nonumber
\\
\delta\Phi^{\alpha(k)\dot\alpha(k)} &=&
D\xi^{\alpha(k)\dot\alpha(k)}
+ k e^\alpha{}_{\dot\alpha} \eta^{\alpha(k-1)\dot\alpha(k+1)}
+ k e_\alpha{}^{\dot\alpha} \eta^{\alpha(k+1)\dot\alpha(k-1)}
\nonumber
\\
 && + \frac{\mu_{k}}{k(k-1)}k^2 e^{\alpha\dot\alpha}
\xi^{\alpha(k-1)\dot\alpha(k-1)}
+ \mu_{k+1} e_{\alpha\dot\alpha} \xi^{\alpha(k+1)\dot\alpha(k+1)},
\nonumber
\\
\delta\Omega^{\alpha(2)} &=&
D\eta^{\alpha(2)} + 2\beta_{2} e^\alpha{}_{\dot\alpha} 
\xi^{\alpha\dot\alpha} + 3\mu_{2} e_{\alpha\dot\alpha} 
\eta^{\alpha(3)\dot\alpha}, 
\\
\delta\Phi^{\alpha\dot\alpha} &=&
D\xi^{\alpha\dot\alpha} + e^\alpha{}_{\dot\alpha} \eta^{\dot\alpha(2)}
+ e_\alpha{}^{\dot\alpha} \eta^{\alpha(2)}
+ \frac{\mu_1}{2} e^{\alpha\dot\alpha} \xi
+ \mu_{2} e_{\alpha\dot\alpha} \xi^{\alpha(2)\dot\alpha(2)}, \nonumber
\\
\delta B^{\alpha(2)} &=& \eta^{\alpha(2)}, \qquad
\delta A = D\xi + \mu_1 e_{\alpha\dot\alpha} \xi^{\alpha\dot\alpha},
\nonumber
\\
\delta \pi^{\alpha\dot\alpha} &=& \xi^{\alpha\dot\alpha}, \qquad
\delta \phi = \xi. \nonumber
\end{eqnarray}
Here, in case of $k=s-1$, the terms which contain the fields with more
than $2s-2$ indices should be omitted. The gauge invariance condition
leads to the following recurrent relations for $\mu_k$, $\beta_k$:
\begin{eqnarray}
\frac{(k+2)}{k}\mu_{k+1}{}^2 &=& \frac{(k+1)}{(k-1)}\mu_k{}^2
- 2\beta_{k+1}(k+1) + 2\lambda^2(k+1) \nonumber,
\\
3 \mu_2{}^2 &=& \mu_1{}^2 - 4\beta_2 + 4\lambda^2,
\\
(k-1)k\beta_k &=& (k+2)(k+1)\beta_{k+1}. \nonumber
\end{eqnarray}

One can see that the general solution of these relations depends on
two free parameters. In case of the finite spin $s$, the condition
$\mu_s=0$ reduces the number of free parameters to one. We use the
"mass" parameter 
$$
M^2=\frac{s(s-1)\mu_{s-1}^2}{2(s-2)}
$$
as the second one. We put the word "mass" in quote since the
translation generators of the (A)dS space do not commute, and the
square of momentum $P^2$ is not thus a Casimir operator anymore.
Hence, there is no straightforward generalization of the notion of the
mass to the constant curvature space. However, in case of completely
symmetric fields (which is the only type of fields we need in four
dimensional case) we may propose a consistent definition for the
massless limit. Namely, this is  the limit where the main gauge field
(i.e. that described by $\Phi^{\alpha(s-1)\dot\alpha(s-1)}$, 
$\Omega^{\alpha(s)\dot\alpha(s-2)}$, 
$\Omega^{\alpha(s-2)\dot{\alpha}(s)}$) decouples from all the
Stueckelberg ones. This corresponds to the limit $\mu_{s-1} \to 0$
when the Lagrangian splits in two independent parts, one containing
$\Phi^{\alpha(s-1)\dot\alpha(s-1)}$, 
$\Omega^{\alpha(s)\dot\alpha(s-2)}$, 
$\Omega^{\alpha(s-2)\dot\alpha(s)}$, while the other one --- the rest
of the fields. As for the concrete normalization, we choose it so that
this parameter coincides with the usual mass in the flat limit 
$\lambda \to 0$. The coefficients $\mu_k$, $\beta_k$ parametrized by
$s$ and $M$ have the following form:
\begin{eqnarray}
\mu_{k}{}^2 &=&
\frac{(s-k)(s+k+1)(k-1)}{k(k+1)^2}
\big[M^2 + (s+k)(s-k-1)\lambda^2 \big], \nonumber
\\
\mu_{1}{}^2 &=& \frac{(s-1)(s+2)}{2}
\big[M^2 + (s+1)(s-2)\lambda^2 \big], 
\\
\beta_k &=& \frac{s(s+1)}{(k-1)k^2(k+1)} 
\big[M^2 + s(s-1)\lambda^2 \big]. \nonumber
\end{eqnarray}
In case of the infinite spin, the notions of spin is inapplicable; we
choose the lowest coefficients $\beta_2$ and $\mu_1$ as the two free
parameters:
\begin{eqnarray}
\mu_{k}{}^2 &=& \frac{(k-1)}{(k+1)}\big[
\mu_1{}^2-\frac{6(k-1)(k+2)}{k(k+1)}\beta_2 + (k-1)(k+2)\lambda^2
\big], \nonumber
\\
\beta_k &=& \frac{12\beta_2}{(k-1)k^2(k+1)}.
\end{eqnarray}

Now let us discuss the hermiticity of the Lagrangian. In case of the
finite spin it implies that all $\mu_k{}^2$ are non-negative. For flat
and $AdS$ spaces that leads to the condition $M^2\ge 0$. In case of
equality $M^2=0$ in $AdS$, the highest field decouples while in flat
space the Lagrangian splits into $s+1$ massless ones. In $dS$ space
($\lambda^2 < 0$) the hermiticity condition leads to the appearance
of the so-called unitary forbidden region $M^2 < -s(s-1)\lambda^2$. 
At the boundary of this region the spin-0 component decouples and we
obtain the first partially massless limit. Inside the forbidden region
we obtain a number of other partially massless ones. Indeed, the
Lagrangian splits into two independent parts at the values of mass
corresponding to the condition $\mu_{k-1}=0$:
$$
M^2 = - (s+k-1)(s-k) \lambda^2. 
$$
In this case, one of the two parts contains the components with spins
$\overline{k,s}$, while the other contains components
$\overline{0,k-1}$; only one of the two parts is unitary. In $dS$, it
is the part with components $\overline{k,s}$ which is unitary; the
lower fields entering the non-unitary part decouple. The resulting
theory is hence unitary, even though the value of the mass lays in the
unitary forbidden region.  

In case of infinite spin, it is convenient to introduce a variable
$y_k=k^2+k-2$. Then, the sign of $\mu_k{}^2$ is determined by a square
trinomial on $y_k$:
$$
\mu_{k}{}^2 \propto \mu_1{}^2(y_k+2)-6y_k\beta_2+y_k(y_k+2)\lambda^2.
$$
It immediately follows that in $dS$ space no infinite spin particle
can exist, since all the $\mu_k{}^2$, starting from sufficiently large
$k$ are negative. Consider the flat case first. There exists a whole
set of unitary solutions with the complete spectrum of helicities
$\overline{0, \pm\infty}$. The unitarity condition reads:
\begin{equation}
\mu_1{}^2 \ge 0, \qquad \mu_1{}^2 > 6\beta_2.
\end{equation}
Most of these solutions are tachyonic, while the solution 
$\mu_1{}^2 = 6\beta_2$ corresponds to the massless infinite spin
field. Note, that it is this solution that can be obtained from the
massive finite spin one if one takes the limit $M \to 0$, $s \to
\infty$ so that $Ms = const$.

Besides, for $\mu_1{}^2 < 0$ the situation analogous to the 
partially massless limit in $dS$ is possible. Indeed, if we set 
$$
\mu_{s-1} = 0 \quad \Rightarrow \quad \mu_1{}^2(y_s+2) = 6\beta_s y_s,
$$
then the non-unitary part decouples so that the components
$\overline{s,+\infty}$ form an unitary theory. 

The $AdS$ case is more complicated. Here we also have a whole set of
solutions with the complete spectrum of helicities 
$\overline{0,\pm\infty}$. The unitarity region is an infinite area
with piece-wise linear boundary (in coordinates $\mu_1{}^2$, 
$\beta_2$):
\begin{eqnarray}
12\beta_2 &\in&
\big[(k-1)k^2(k+1)\lambda^2;k(k+1)^2(k+2)\lambda^2\big], \qquad
k\in\mathbb{N}, \nonumber
\\
\mu_1{}^2 &>& (k^2+k-2)\big[\frac{6\beta_2}{(k^2+k)} - \lambda^2\big].
\end{eqnarray}
Each segment of the boundary corresponds to the condition
$\mu_k{}^2>0$. Once again, a set of partially massless limits is also
possible. The solution for partially massless limit corresponding to
the components $\overline{s,+\infty}$ can be written in a form very
similar to that of the massive finite spin case:
\begin{eqnarray}
\beta_k &=& \frac{s(s+1)\hat{M}^2}{(k-1)k^2(k+1)}, \qquad
\hat{M}^2 < s(s+1)\lambda^2, \nonumber
\\
\mu_k{}^2 &=& \frac{(k-s)(k+s+1)}{k(k+1)} [k(k+1)\lambda^2 - 
\hat{M}^2].
\end{eqnarray}

\subsection{Gauge invariant curvatures}

As it was already mentioned, one of the advantages of the frame-like
formalism is the possibility to construct a complete set of
gauge invariant objects, or curvatures. However, in contrast to the
case of massless spin-2 particle, for the massless spin $s > 2$
particles one has to introduce the so-called extra fields, which do
not enter the free Lagrangian. They, however, do transform under the
gauge transformations and enter the curvatures as well as the
interaction Lagrangian. In the massless case the complete set of
fields is $\Omega^{\alpha(s-1+m)\dot\alpha(s-1-m)}$, $|m|\le s-1$,
where the field with $m=0$ is the physical one, while the fields with
$m= \pm 1$ are the auxiliary ones. Thus in the massive case we need
the following set of one-forms \cite{PV10}
$\Omega^{\alpha(k+m)\dot\alpha(k-m)}$, $|m|\le k\le s-1$.
However, our Lagrangian contains zero-forms as well, and it appears
that to construct the complete set of the gauge invariant objects one
has to introduce the following set of zero-forms 
$W^{\alpha(k+m)\dot\alpha(k-m)}$, $m\le k\le s-1$, so that we have a
one to one correspondence between the one-forms and the zero-forms
(seee Figure 2a). 
\begin{figure}[htb]
\begin{center}
\begin{picture}(260,120)

\put(10,10){\vector(0,1){100}}
\put(0,100){\makebox(10,10){$m$}}
\put(10,60){\vector(1,0){105}}
\put(105,50){\makebox(10,10){$k$}}

\put(10,60){\circle*{2}}
\put(10,60){\circle{4}}

\multiput(20,50)(0,10){3}{\circle*{2}}
\multiput(20,50)(0,10){3}{\circle{4}}

\multiput(30,40)(0,10){5}{\circle*{2}}
\multiput(30,40)(0,10){5}{\circle{4}}

\multiput(40,30)(0,10){7}{\circle*{2}}
\multiput(40,30)(0,10){7}{\circle{4}}

\multiput(50,20)(0,10){9}{\circle*{2}}
\multiput(50,20)(0,10){9}{\circle{4}}

\multiput(60,110)(10,0){5}{\oval[0.5](4,4)}
\multiput(60,100)(10,0){5}{\oval[0.5](4,4)}
\multiput(60,90)(10,0){5}{\oval[0.5](4,4)}
\multiput(60,80)(10,0){5}{\oval[0.5](4,4)}
\multiput(60,70)(10,0){5}{\oval[0.5](4,4)}
\multiput(60,60)(10,0){5}{\oval[0.5](4,4)}
\multiput(60,50)(10,0){5}{\oval[0.5](4,4)}
\multiput(60,40)(10,0){5}{\oval[0.5](4,4)}
\multiput(60,30)(10,0){5}{\oval[0.5](4,4)}
\multiput(60,20)(10,0){5}{\oval[0.5](4,4)}
\multiput(60,10)(10,0){5}{\oval[0.5](4,4)}


\put(150,10){\vector(0,1){100}}
\put(140,100){\makebox(10,10){$m$}}
\put(150,60){\vector(1,0){105}}
\put(245,50){\makebox(10,10){$k$}}

\multiput(170,40)(0,10){5}{\circle*{2}}

\multiput(180,30)(0,10){7}{\circle*{2}}
\put(180,30){\circle{4}}
\put(180,90){\circle{4}}

\multiput(190,20)(0,10){9}{\circle*{2}}
\multiput(190,90)(0,10){2}{\circle{4}}
\multiput(190,20)(0,10){2}{\circle{4}}

\multiput(200,110)(10,0){5}{\oval[0.5](4,4)}
\multiput(200,100)(10,0){5}{\oval[0.5](4,4)}
\multiput(200,90)(10,0){5}{\oval[0.5](4,4)}

\multiput(200,30)(10,0){5}{\oval[0.5](4,4)}
\multiput(200,20)(10,0){5}{\oval[0.5](4,4)}
\multiput(200,10)(10,0){5}{\oval[0.5](4,4)}

\end{picture}
\end{center}
\caption{a) Left figure shows the spectrum of one-forms (dots),
Stueckelberg (circles) and gauge invarinat (squares) zero-forms for
the massive boson with $s=5$. b) Right figure --- partially massless
case with $n=3$. Dots without surrounding circles correspond to the
fields of Skvortsov-Vasiliev formalism.}
\end{figure}
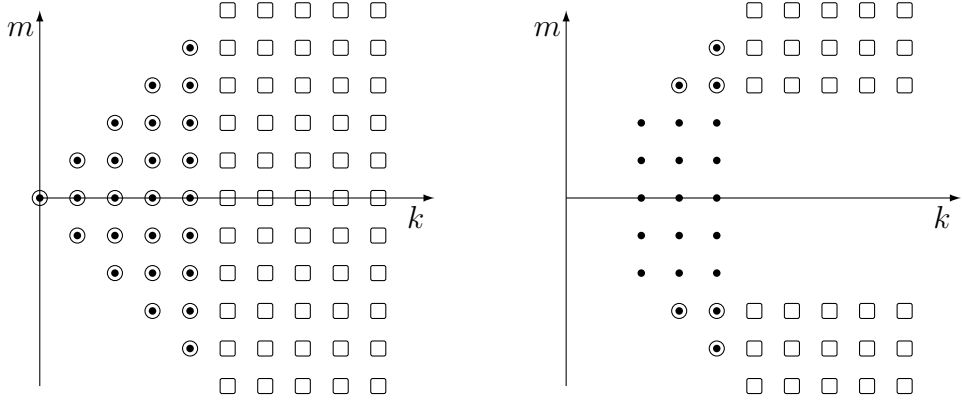
From here on in the section, the notations are unified: 
$\Phi^{\alpha(k)\dot\alpha(k)}\equiv\Omega^{\alpha(k)\dot\alpha(k)}$,
$B^{\alpha(2)} \equiv W^{\alpha(2)}$, $\pi^{\alpha\dot\alpha} \equiv
W^{\alpha\dot\alpha}$, $\phi\equiv W$. The gauge transformation law of
the physical and auxiliary fields has already been obtained. Then the
most general ansatz for the extra fields gauge transformation, up to
normalization choice, is (see Appendix B about the coefficients 
$\alpha^{ij}_{k,m}$):
\begin{eqnarray}
\delta\Omega^{\alpha(k+m){\dot{\alpha}}(k-m)} &=&
D\eta^{\alpha(k+m)\dot\alpha(k-m)}
+ (k+m)(k-m)\alpha^{--}_{k,m} e^{\alpha\dot\alpha}
\eta^{\alpha(k+m-1)\dot\alpha(k-m-1)} \nonumber
\\
 && + \alpha^{++}_{k,m} e_{\alpha\dot\alpha}
\eta^{\alpha(k+m+1)\dot\alpha(k-m+1)}
+ (k+m)\alpha^{-+}_{k,m} e^\alpha{}_{\dot\alpha}
\eta^{\alpha(k+m-1)\dot\alpha(k-m+1)} \nonumber
\\
 && +(k-m) e_\alpha{}^{\dot\alpha}
\eta^{\alpha(k+m+1)\dot\alpha(k-m-1)},
\\
\delta\Omega^{\alpha(2k)} &=&
D\eta^{\alpha(2k)} + \alpha^{++}_{k,k} e_{\alpha\dot\alpha}
\eta^{\alpha(2k+1)\dot\alpha} +2 k\alpha^{-+}_{k,k} 
e^\alpha{}_{\dot\alpha} \eta^{\alpha(2k-1)\dot\alpha}, \nonumber
\\
\delta W^{\alpha(k+m)\dot\alpha(k-m)} &=&
\eta^{\alpha(k+m)\dot\alpha(k-m)}, \nonumber
\end{eqnarray}
where
\begin{eqnarray*}
\alpha_{k,1}^{++} &=& \frac{(k+2)}{k}\mu_{k+1}, \qquad
\alpha_{k,1}^{-+} = \beta_{k+1}, \qquad
\alpha_{k,1}^{--} = \frac{\mu_k}{(k-1)(k+2)} 
\\
\alpha_{k,0}^{++} &=& \mu_{k+1}, \qquad
\alpha_{k,0}^{-+} = 1, \qquad
\alpha_{k,0}^{--} = \frac{\mu_k}{k(k-1)}
\end{eqnarray*}
The gauge transformations completely define the form of the
curvatures. For $k\ge2$ those curvatures are: 
\begin{eqnarray}
R^{\alpha(k+m)\dot\alpha(k-m)} &=&
D\Omega^{{\alpha(k+m)\dot\alpha(k-m)}}
+ (k+m)(k-m)\alpha^{--}_{k,m} e^{\alpha\dot\alpha}
\Omega^{\alpha(k+m-1)\dot\alpha(k-m-1)} \nonumber
\\
 && + \alpha^{++}_{k,m} e_{\alpha\dot\alpha}
\Omega^{\alpha(k+m+1)\dot\alpha(k-m+1)}
+ (k+m)\alpha^{-+}_{k,m} e^\alpha{}_{\dot\alpha}
\Omega^{\alpha(k+m-1)\dot\alpha(k-m+1)} \nonumber
\\
 && + (k-m) e_\alpha{}^{\dot\alpha}
\Omega^{\alpha(k+m+1)\dot\alpha(k-m-1)},
\qquad\qquad\qquad\qquad (0\le m<k), \nonumber
\\
R^{\alpha(2k)} &=& D\Omega^{\alpha(2k)}
+ \alpha^{++}_{k,m} e_{\alpha\dot\alpha}
\Omega^{\alpha(2k+1)\dot\alpha}
+ 2k\alpha^{-+}_{k,k} e^\alpha{}_{\dot\alpha}
\Omega^{\alpha(2k-1)\dot\alpha} \nonumber
\\
 && - 4k(2k-1)\alpha^{-+}_{k,k}\alpha^{--}_{k,k-1} E^{\alpha(2)}
W^{\alpha(2k-2)} - 2\alpha^{++}_{k,k} E_{\alpha(2)} W^{\alpha(2k+2)}
\nonumber
\\
 && - \frac{\alpha^{-+}_{k+1}}{k+1} E^\alpha{}_\beta 
W^{\alpha(2k-1)\beta},
\\
C^{\alpha(k+m)\dot\alpha(k-m)} &=&
DW^{\alpha(k+m)\dot\alpha(k-m)} - \Omega^{\alpha(k+m)\dot\alpha(k-m)}
\nonumber
\\
 && + (k+m)(k-m)\alpha^{--}_{k,m} e^{\alpha\dot\alpha}
W^{\alpha(k+m-1)\dot\alpha(k-m-1)} \nonumber
\\
 && + \alpha^{++}_{k,m} e_{\alpha\dot\alpha}
W^{\alpha(k+m+1)\dot\alpha(k-m+1)}
+ (k-m) e_\alpha{}^{\dot\alpha}
W^{\alpha(k+m+1)\dot\alpha(k-m-1)} \nonumber
\\
 && + (k+m)\alpha^{-+}_{k,m} e^\alpha{}_{\dot\alpha}
W^{\alpha(k+m-1)\dot\alpha(k-m+1)}. \nonumber
\end{eqnarray}
The expressions for the lower spin curvatures have different
coefficients and thus have to be written out separately:
\begin{eqnarray}
R^{\alpha\dot\alpha} &=& D\Omega^{\alpha\dot\alpha}
+ e^\alpha{}_{\dot\alpha} \Omega^{\dot\alpha(2)}
+ e_\alpha{}^{\dot\alpha} \Omega^{\alpha(2)}
+ \frac{\mu_1}{2} e^{\alpha\dot\alpha} \Omega
+ \mu_{2} e_{\alpha\dot\alpha} \Omega^{\alpha(2)\dot\alpha(2)},
\nonumber
\\
R^{\alpha(2)} &=& D\Omega^{\alpha(2)}
+ \mu_0^2 e^\alpha{}_{\dot\alpha} \Omega^{\alpha\dot\alpha}
+ 3\mu_2 e_{\alpha\dot\alpha} \Omega^{\alpha(3)\dot\alpha} \nonumber
\\
 && - \mu_1^2 E^\alpha{}_\beta W^{\alpha\beta}
- \mu_0^2\mu_1 E^{\alpha(2)} W 
- 6\mu_2 E_{\alpha(2)} W^{\alpha(4)}, \nonumber
\\
C^{\alpha\dot\alpha} &=& DW^{\alpha\dot\alpha}
- \Omega^{\alpha\dot\alpha}
+ e^\alpha{}_{\dot\beta} W^{\dot\alpha\dot\beta}
+ e_\beta{}^{\dot\alpha} W^{\alpha\beta}
+ \frac{\mu_1}{2} e^{\alpha\dot\alpha} W
+ \mu_2 e_{\alpha\dot\alpha} W^{\alpha(2)\dot\alpha(2)}, 
\\
C^{\alpha(2)} &=& DW^{\alpha(2)}
- \Omega^{\alpha(2)}
+ \mu_0^2 e^\alpha{}_{\dot\alpha} W^{\alpha\dot\alpha}
+ 3\mu_2 e_{\alpha\dot\alpha} W^{\alpha(3)\dot\alpha}, \nonumber
\\
 R &=& D\Omega
+ \mu_1 e_{\alpha\dot\alpha} \Omega^{\alpha\dot\alpha} - 2\mu_1
E_{\alpha(2)} W^{\alpha(2)}
- 2\mu_1 E_{\dot\alpha(2)} W^{\dot\alpha(2)}, \nonumber
\\
C &=& DW - \Omega
+ \mu_1 e_{\alpha\dot\alpha} W^{\alpha\dot\alpha}. \nonumber
\end{eqnarray}

It is convenient to introduce auxiliary coefficients
$\alpha^{++}_{k},\alpha^{--}_{k},\alpha^{-+}_{m}$ which have one index
only. The coefficients $\alpha^{ij}_{k,m}$ can be expressed
in terms of these auxiliary ones as follows:
\begin{eqnarray}
\alpha^{++}_{k,m} &=& \frac{\alpha^{++}_{k}}{(k-m+1)(k-m+2)},
\nonumber
\\
\alpha^{--}_{k,m} &=& \frac{\alpha^{--}_{k}}{(k+m)(k+m+1)}, 
\\
\alpha^{-+}_{k,m} &=& \frac{\alpha^{-+}_{m}}.
{(k-m+1)(k-m+2)(k+m)(k+m+1)} \nonumber
\end{eqnarray}
These expressions are applicable both for the finite and the infinite
spin cases. To express the auxiliary coefficients, we use the same
parameter choice as in the previous subsection. Namely, spin $s$ and
"mass" parameter $M$ in case of the finite spin:
\begin{eqnarray}
\alpha^{++}_{k-1}{}^2 &=& k(k-1)(s-k)(s+k+1)
\big[M^2 + (s+k)(s-k-1)\lambda^2\big], \nonumber
\\
\alpha^{--}_{k}{}^2 &=& \frac{(s-k)(s+k+1)}{k(k-1)}
\big[M^2 + (s+k)(s-k-1)\lambda^2\big],
\\
\alpha^{-+}_{m} &=& (s-m+1)(s+m)\big[M^2 + (s-m)(s+m-1)\lambda^2\big],
\nonumber
\end{eqnarray}
and the lowest coefficients $\mu_1$, $\beta_2$ in case of the infinite
spin:
\begin{eqnarray}
\alpha^{++}_{k-1}{}^2 &=& k(k-1) \big[
\mu_1{}^2k(k+1) - 6(k-1)(k+2)\beta_2 + (k-1)k(k+1)(k+2)\lambda^2
\big],
\nonumber
\\
\alpha^{--}_{k}{}^2 &=& \frac{1}{k(k-1)} \big[
\mu_1{}^2k(k+1) - 6(k-1)(k+2)\beta_2 + (k-1)k(k+1)(k+2)\lambda^2
\big],
\\
\alpha^{-+}_{k} &=& \big[
\mu_1{}^2(k-1)k-6(k-2)(k+1)\beta_2+(k-2)(k-1)k(k+1)\lambda^2 \big].
\nonumber
\end{eqnarray}
Note the useful relation
$\alpha^{--}_{m-1}\alpha^{++}_{m-2}=\alpha^{-+}_{m}$. The relation is
a general rule, i.e. it holds not only for the bosonic
$\alpha^{ij}_{n}$, but for their fermionic analogues as well.

Note that the hermiticity of the curvatures 
$\big(R^{\alpha(k+m)\dot\alpha(k-m)}\big)^\dagger=
R^{\alpha(k-m)\dot\alpha(k+m)}$ requires the coefficients 
$\alpha^{++}_{k,m}{}^2$, $\alpha^{--}_{k,m}{}^2$ to be real. One can
see that
$\alpha^{--}_{k,m}{}^2\propto\alpha^{++}_{k-1,m}{}^2\propto\mu_k^2$,
$\alpha^{-+}_{m+1}{}^2\propto\mu_m^2$. Hence, the hermiticity of the
Lagrangian is equivalent to the hermiticity of the curvatures. 

In case of the partially massless limit where the unitary part
contains the components $\overline{k,s}$, all the lower spin fields
(i.e. $\Omega^{\alpha(l+m)\dot\alpha(l-m)}$,
$W^{\alpha(l+m)\dot\alpha(l-m)}$ for $l<k-1$) completely decouple.
Besides, all the zero-forms $W^{\alpha(l+m)\dot\alpha(l-m)}$ with 
$l \ge k-1$, $|m| \le k-1$ also decouple. This leaves us with the set
of one forms $\Omega^{\alpha(l+m)\dot\alpha(l-m)}$ with $l \ge k-1$, 
$|m| \le k-1$ (which exactly correspond to the Skvortsov-Vasiliev
formalism \cite{SV06}, see below) as well as the pairs of one-forms
and zero-forms with $l \ge k$, $l \ge |m| \ge k$ (see Figure 2b).

\subsection{Lagrangian in terms of the curvatures}

The existence of the complete set of gauge invariant curvatures
allows us to rewrite the Lagrangian in the explicitly gauge invariant
form. The most general ansatz for the Lagrangian in terms of the
curvatures is:
\begin{eqnarray}
\label{bcLag}
-i\mathcal{L} &=&
\sum_{k=0}^{s-1}\sum_{m=-k}^k (-1)^{k+1} a_{k,m}
R^{\alpha(k+m)\dot\alpha(k-m)} R_{\alpha(k+m)\dot\alpha(k-m)}
\nonumber
\\
 && + \sum_{k=0}^{s-2}\sum_{m=-k}^k (-1)^{k+1}b_{k,m}
R^{\alpha(k+m)\dot\alpha(k-m)} e^{\alpha\dot\alpha}
C_{\alpha(k+m+1)\dot\alpha(k-m+1)} \nonumber
\\
 && + \sum_{k=1}^{s-1}\sum_{m=-k+1}^{k-1} (-1)^{k+1}c_{k,m}
R^{\alpha(k+m)\dot\alpha(k-m)} e_{\alpha\dot\alpha}
C_{\alpha(k+m-1)\dot\alpha(k-m-1)} \nonumber
\\
 && + \sum_{k=1}^{s-1}\sum_{m=-k}^{k-1} (-1)^{k+1}d_{k,m}
R^{\alpha(k+m)\dot\alpha(k-m)} e^\beta{}_{\dot\alpha}
C_{\alpha(k+m)\beta\dot\alpha(k-m-1)} \nonumber
\\
 && - \sum_{k=1}^{s-1}\sum_{m=-k+1}^{k} (-1)^{k+1}d_{k,-m}
R^{\alpha(k+m)\dot\alpha(k-m)} e_\alpha{}^{\dot\beta}
C_{\alpha(k+m-1)\dot\beta\dot\alpha(k-m)} \nonumber
\\
 && + \sum_{k=1}^{s-1}\sum_{m=-k+1}^k (-1)^{k+1}e_{k,m}
C^{\alpha(k+m)\dot\alpha(k-m)} E^\beta{}_\alpha
C_{\alpha(k+m-1)\beta\dot\alpha(k-m)} \nonumber
\\
 && - \sum_{k=1}^{s-1}\sum_{m=-k}^{k-1} (-1)^{k+1}e_{k,-m}
C^{\alpha(k+m)\dot\alpha(k-m)} E^{\dot\beta}{}_{\dot\alpha}
C_{\alpha(k+m)\dot\beta\dot\alpha(k-m-1)} \nonumber
\\
 && + \sum_{k=0}^{s-2}\sum_{m=-k}^k (-1)^{k+1}f_{k,m}
C^{\alpha(k+m)\dot\alpha(k-m)} E^{\alpha(2)}
C_{\alpha(k+m+2)\dot\alpha(k-m)} \nonumber
\\
 && - \sum_{k=0}^{s-2}\sum_{m=-k}^k (-1)^{k+1}f_{k,-m}
C^{\alpha(k+m)\dot\alpha(k-m)} E^{\dot\alpha(2)}
C_{\alpha(k+m)\dot\alpha(k-m+2)},
\end{eqnarray}
where $a_{k,m}=-a_{k,-m}$, $b_{k,m}=-b_{k,-m}$, $c_{k,m}=-c_{k,-m}$
(for the hermiticity of the Lagrangian). The most straightforward way
to calculate the coefficients $a_{k,m}-f_{k,m}$ is to substitute the
curvatures with their expressions via fields and require the result to
be equal to (\ref{bLag}). It is much more convenient, however, to
require the Lagrangian equations to match. Since the equations are
gauge invariant, they can be expressed via the curvatures as well.
Hence, the curvatures can be used during the whole process of
calculation reducing the number of terms. The requirement of matching
the equations is equivalent to the extra field decoupling conditions:
\begin{eqnarray}
\frac{\delta\mathcal{L}}
{\delta\Omega^{\alpha(k-1-m)\dot\alpha(s-1+m)}} &=& 0, \qquad 
|m| \ge 2, \nonumber
\\
\frac{\delta\mathcal{L}}
{\delta W^{\alpha(k-1-m)\dot\alpha(k-1+m)}} &=& 0, \qquad k \ge 2,
\end{eqnarray}
up to the normalization, which is fixed by the normalization of the
equations for the physical and auxiliary fields:
\begin{eqnarray}
&& \frac{\delta \mathcal{L}}
{\delta\Omega^{\alpha(k-1)\dot\alpha(k+1)}} = 2(-1)^{k+1} 
e^\beta{}_{\dot\gamma} R_{\alpha(k-1)\beta\dot\alpha(k)},
\nonumber 
\\
&& \frac{\delta \mathcal{L}}{\delta W^{\alpha(2)}} = - 2\mu_1
E_{\alpha(2)} R, \qquad
\frac{\delta \mathcal{L}}{\delta W^{\alpha\dot\alpha}}
= - 24\mu_1\beta_2 e_{\alpha\dot\alpha} C ,\nonumber
\\
&& \frac{\delta \mathcal{L}}
{\delta\Omega^{\alpha(k)\dot\alpha(k)}} = 2(-1)^{k+1} 
e_\alpha{}^{\dot\gamma} R_{\alpha(k-1)\dot\gamma\dot\alpha(k)}
+ h.c.,
\\
&& \frac{\delta \mathcal{L}}{\delta \Omega} = 2 \mu_1 E_{\alpha(2)}
C^{\alpha(2)} + h.c., \qquad  
\frac{\delta \mathcal{L}}{\delta W} = - 24\mu_1\beta_2 
e_{\alpha\dot\alpha} C^{\alpha\dot\alpha}. \nonumber
\end{eqnarray}

Those conditions yield a system of linear equations for
$a_{k,m}-f_{k,m}$. However, there is an arbitrarity in the choice of
$a_{k,m}-f_{k,m}$. It stems from the fact that there exist terms
quadratic in the curvatures that are equal to the total derivative of
some object, which does not alter the equations of motion (see
Appendix B):
\begin{eqnarray}
i(\mathcal{L}-\mathcal{L}_0) &=&
\sum_{k=0}^{s-1}\sum_{m=1}^k (-1)^{k+1} 
p_{k,m} D( R^{\alpha(k+m)\dot\alpha(k-m)}
C_{\alpha(k+m)\dot\alpha(k-m)}-h.c. ) \nonumber
\\
 && + \sum_{k=1}^{s-1}\sum_{m=0}^{k-1} (-1)^{k+1} 
q_{k,m} D( C^{\alpha(k+m)\dot\alpha(k-m)}
e^\alpha{}_{\dot\alpha} C_{\alpha(k+m+1)\dot\alpha(k-m-1)}-h.c.)
\nonumber
\\
 && + \sum_{k=0}^{s-1}\sum_{m=1}^k (-1)^{k+1} 
r_{k,m} D( C^{\alpha(k+m)\dot\alpha(k-m)} e^{\alpha\dot\alpha}
C_{\alpha(k+m+1)\dot\alpha(k-m+1)}-h.c. ). 
\end{eqnarray}
Hence, the parameters of the Lagrangian is determined up to the shifts
with $p_{k,m}$, $q_{k,m}$ and $r_{k,m}$ (see their explicit
expressions in Appendix B). By an appropriate choice of
$p_{k,m},q_{k,m},r_{k,m}$ one can set to zero all the $b_{k,m}$,
$c_{k,m}$  and $d_{k,m}$ for $m\ge 0$. It follows from the equations
that all the $d_{k,m}$, $e_{k,m}$,  $f_{k,m}$, except $e_{k,k}$,
$f_{k,k}$ turn out to be zero as well. We obtain the following
expressions for the remaining coefficients $a_{k,m}^{(0)}$, 
$e_{k,k}^{(0)}$, $f_{k,k}^{(0)}$:
\begin{eqnarray}
a_{k,\pm m}^{(0)} &=& \pm
\frac{(k-1)!(k+m+1)!k!}{(k-m)!^2(k-m+1)!\prod_{i=1}^{m}
\alpha^{-+}_{i}}, \qquad m>0 ,\nonumber
\\
e_{k,k}^{(0)} &=& \frac{a_{k,k}\alpha^{-+}_{k+1}}{k+1}, \qquad
f_{k,k}^{(0)} = - 4\alpha^{++}_{k,k}a_{k,k}, \quad k>0,
\\
f_{0,0}^{(0)} &=& - 2\mu_1. \nonumber
\end{eqnarray}
For such choice of the coefficients, the structure of the Lagrangian
simplifies to:
\begin{eqnarray}
-i\mathcal{L} &=&
\sum_{k=0}^{s-1}(-1)^{k+1} \sum_{m=0}^k  a_{k,m}
R^{\alpha(k+m)\dot\alpha(k-m)} R_{\alpha(k+m)\dot\alpha(k-m)}
\nonumber
\\
 && + \sum_{k=1}^{s-1}(-1)^{k+1}e_{k,k}
C^{\alpha(2k)} E^\beta_\alpha C_{\alpha(2k-1)\beta}-h.c. \nonumber
\\
 && + \sum\limits_{k=0}^{s-2} (-1)^{k+1}f_{k,k}
C^{\alpha(2k)} E^{\alpha(2)} C_{\alpha(2k+2)} -h.c.
\end{eqnarray}
Note that the structure of the expression is the same as in
\cite{PV10}. 

One can see that the expression contain singularities in case of
partially massless limits (i.e. for 
$\alpha^{-+}_n=\alpha^{--}_{n-1}=0$). In this case, our ansatz fails.
However, we can return back to the general solution and use the
shifts $p_{k,m},q_{k,m},r_{k,m}$ to remove the poles, so that the
limit $\alpha^{-+}_n\to0$ can be taken. We do this in the most
straightforward way - we set all the singular coefficients 
$a_{k,m}^{(0)}$, $m\ge n$, $e_{k,k}^{(0)}$, $f_{k,k}^{(0)}$, $k>n$ to
zero, while preserving zero values of $b_{k,m}$, $c_{k,m}$ and
$d_{k,m}$ ($m\ne n-1$). Then, the coefficients with $k<n$ remain 
the same, except the coefficients $e_{n-1,n-1}$, $f_{n-2,n-2}$,
$f_{n-1,n-1}$, which become zero. The non-zero coefficients for 
$k\ge n$ are:
\begin{eqnarray}
\pm a_{k,\pm m} &=& \frac{(k-1)!(k+m+1)!k!}{(k-m)!^2(k-m+1)!
\prod_{i=1}^{m}\alpha^{-+}_{i}}, \qquad 0<m<n, \nonumber
\\
d_{k,n-1} &=& -2(k-n+1)a_{k,n-1}, 
\\
e_{k,n} &=& -(k-n+1)(k-n+2)a_{k,n-1}, \nonumber
\\
e_{k,-n} &=& -(k-n+1)(k-n)a_{k,n-1}. \nonumber
\end{eqnarray}
The Lagrangian has the structure:
\begin{eqnarray}
\label{bcLagPML}
-i\mathcal{L} &=& -i\mathcal{L}^{(0,n-2)} +
\sum_{k=n-1}^{s-1}\sum_{m=-n+1}^{n-1} (-1)^{k+1} a_{k,m}
R^{\alpha(k+m)\dot\alpha(k-m)} R_{\alpha(k+m)\dot\alpha(k-m)}
\nonumber
\\
 && + \sum_{k=n}^{s-1} (-1)^{k+1}d_{k,n-1}\big[
R^{\alpha(k+n-1)\dot\alpha(k-n+1)} e^\beta{}_{\dot\alpha}
C_{\alpha(k+n-1)\beta\dot\alpha(k-n)}- h.c. \big] \nonumber
\\
 && + \sum\limits_{k=n}^{s-1} (-1)^{k+1}e_{k,n}
\big[ C^{\alpha(k+n)\dot\alpha(k-n)} E^\beta{}_\alpha
C_{\alpha(k+n-1)\beta\dot\alpha(k-n)} - h.c. \big] \nonumber
\\
 && + \sum_{k=n}^{s-1} (-1)^{k+1}e_{k,-n} \big[
C^{\alpha(k+n)\dot\alpha(k-n)} E^{\dot\beta}{}_{\dot\alpha}
C_{\alpha(k+n-1)\beta\dot\alpha(k-n)} - h.c. \big].
\end{eqnarray}
Here $\mathcal{L}^{(0,n-2)}$ contains all the terms with $k\le n-2$.

One can see that the Lagrangian splits in two parts containing the
fields with $k\ge n-1$ and $k<n-1$ respectively. This is an expected
result for the partially massless limit. Note that the fields
$W^{\alpha(k+m)\dot\alpha(k-m)}$, $k \ge n-1$, $|m|\le n-1$ also do
not enter the Lagrangian for the components $\overline{n,s}$.

\subsection{Unfolded equations}

Let us consider an unfolded formulation for massive spin-$s$ boson.
Using the explicit expressions for the curvatures given above, one can
straightforwardly check that it is consistent to set to zero most of
them, namely:
\begin{eqnarray}
0 &=& R^{\alpha(s-1+m){\dot{\alpha}}(s-1-m)}, \qquad |m| \ne s-1,
\nonumber
\\
0 &=& R^{\alpha(k+m){\dot{\alpha}}(k-m)}, \qquad k < s-1 
\\
0 &=& C^{\alpha(k+m){\dot{\alpha}}(k-m)}, \qquad k < s-1. \nonumber
\end{eqnarray}
As for the remaining curvatures, to write consistent equations for
them one has to introduce a first set of the gauge invariant 
zero-forms:
\begin{eqnarray}
0 &=& R^{\alpha(2s-2)} - 2E_{\alpha(2)} W^{\alpha(2s)}, \nonumber
\\
0 &=& C^{\alpha(s+m-1)\dot\alpha(s-m-1)} +
e_{\alpha\dot\alpha}W^{\alpha(s+m)\dot\alpha(s-m)}.
\end{eqnarray}
These equations connect the gauge sector with the infinite tail
containing gauge-invariant zero-forms only. Indeed, the equations for
these new zero-forms require introduction of additional zero-forms and
so on. This procedure leads to the infinite set of the gauge invariant
zero-forms $W^{\alpha(k+m)\dot\alpha(k-m)}$, $k\ge s,|m| \le s$. Thus
the complete set of one-forms and zero-forms for the massive spin-$s$
boson is equal to the sum of the one-forms and zero-forms necessary
for the unfolded formulation for the massless fields with spins 
$\overline{0,s}$ (see Figure 2a). The main difference is that a part
of zero-forms, namely, $W^{\alpha(k+m)\dot\alpha(k-m)}$, $|m| \le k\le
s$ are not gauge invariant but play the role of the Stueckelberg
fields. The equations for the tail are similar to their massless
analogues; however, just as every other object (Lagrangian, gauge
transformations, curvatures), they have to contain the cross-terms.
The most general ansatz for the tail equations, up to the
normalization choice, is: 
\begin{eqnarray}
0 &=& DW^{\alpha(k+m)\dot\alpha(k-m)}
+ (k+m)(k-m)\beta^{--}_{k,m} e^{\alpha\dot\alpha}
W^{\alpha(k+m-1)\dot\alpha(k-m-1)} \nonumber
\\
 && + e_{\beta\dot\beta} W^{\alpha(k+m)\beta\dot\alpha(k-m)\dot\beta}
+ (k-m)\beta^{+-}_{k,m} e_\beta{}^{\dot\alpha}
W^{\alpha(k+m)\beta\dot\alpha(k-m-1)} \nonumber
\\
 && + (k+m)\beta^{-+}_{k,m} e^\alpha{}_{\dot\beta}
W^{\alpha(k+m-1)\dot\alpha(k-m)\dot\beta},
\end{eqnarray}
where $\beta^{-+}_{k,m}=\beta^{-+}_{k,-m}$ due to hermiticity. The
equations must be consistent with each other as well as with the
gauge sector equations. This leads to the unique possible choice of
the coefficients:
\begin{eqnarray}
\beta^{-+}_{k,m} &=& \frac{\beta^{-+}_{m}}{(k+m)(k+m+1)}, \nonumber
\\
\beta^{+-}_{k,m} &=& \frac{\beta^{+-}_{m}}{(k-m)(k-m+1)}, \nonumber
\\
\beta^{--}_{k,m} &=& \frac{\alpha^{-+}_{k+1}}
{(k+m)(k+m+1)(k-m)(k-m+1)}, 
\\
\beta^{-+}_{m} &=& \frac{\alpha^{-+}_{m}}{(s-m)(s-m+1)}, \quad 
1\le m<s, \qquad \beta_{s}^{-+} = \frac{\alpha^{-+}_s}{2}, \nonumber
\\
\beta^{+-}_{m} &=& (s-m-1)(s-m), \quad 0\le m<s-1, \qquad
\beta^{+-}_{s-1} = 2. \nonumber
\end{eqnarray}

In case of the partially-massless limit given by $\alpha^{-+}_{n}=0$,
the curvatures  $C^{\alpha(k+m)\dot\alpha(k-m)}$, $m<n$ decouple.
This corresponds to the equality $\beta^{-+}_{n}=0$ for the unfolded
equations, which means that the fields
$W^{\alpha(k+m)\dot\alpha(k-m)},m<n$ decouple as well (see Figure 2b).
Hence, only the components with $\overline{n,s}$ remain, which is an
expected result.

\subsection{Application to the Skvortsov-Vasiliev formalism}

In the paper of Skvortsov and Vasiliev \cite{SV06}, an approach for
the description of partially massless particles was proposed, which
use the one-forms only. Consider a $\overline{n,s}$ partially massless
limit. The Skvortsov-Vasiliev formalism corresponds to a partial
gauge fixing, when all the zero-forms
$W^{\alpha(k+m)\dot\alpha(k-m)}$, $k<s$, $|m|\ge n-1$ are set to zero
(see Figure 2b). Then, the one-form curvatures 
$C^{\alpha(k+m)\dot\alpha(k-m)}$, $|m|\ge n-1$ become:
\begin{equation}
C^{\alpha(k+m)\dot\alpha(k-m)} = - 
\Omega^{\alpha(k+m)\dot\alpha(k-m)}.
\end{equation}
Other one-form curvatures, namely $C^{\alpha(k+m)\dot\alpha(k-m)}$, 
$|m|< n-1$, decouple. The curvatures $R^{\alpha(k+m)\dot\alpha(k-m)}$,
$|m|< k$ do not change since they contain no zero-forms. However, it
is convenient to introduce the modified
$\hat{R}^{\alpha(k+n-2)\dot\alpha(k-n+2)}$ curvature as:
\begin{eqnarray}
\hat{R}^{\alpha(k+n-2)\dot\alpha(k-n+2)} &=&
R^{\alpha(k+n-2)\dot\alpha(k-n+2)} + (k-n+2) e_\alpha{}^{\dot\alpha}
C^{\alpha(k+n-1)\dot\alpha(k-n+1)} \nonumber
\\
 &=& D\Omega^{\alpha(k+n-2)\dot\alpha(k-n+2)}
+ \alpha^{++}_{k,n-2} e_{\alpha\dot\alpha}
\Omega^{\alpha(k+n-1)\dot\alpha(k-n+3)} \nonumber
\\
 && + (k+n-2)(k-n+2)\alpha^{--}_{k,n-2} e^{\alpha\dot\alpha}
\Omega^{\alpha(k+n-3)\dot\alpha(k-n+1)} \nonumber
\\
 && + (k+n-2)\alpha^{-+}_{k,n-2} e^\alpha{}_{\dot\alpha}
\Omega^{\alpha(k+n-3)\dot\alpha(k-n+3)}.
\end{eqnarray}
Let $\hat{R}^{\alpha(k+m)\dot\alpha(k-m)}=
R^{\alpha(k+m)\dot\alpha(k-m)}$, $|m|<n-2$ to make the notations
uniform. Then, the new set of curvatures 
${\hat{R}}^{\alpha(k+m)\dot\alpha(k-m)}$ does not contain  
$\Omega^{\alpha(k+m)\dot\alpha(k-m)}$, $|m|\ge n-1$ at all. In this
case, the Lagrangian (\ref{bcLagPML}) can be rewritten purely in terms
of $\hat{R}^{\alpha(k+m)\dot\alpha(k-m)}$:
\begin{equation}
-i\mathcal{L} = \sum_{k=n-1}^{s-1}\sum_{m=-n+1}^{n-1} (-1)^{k+1}
a_{k,m} \hat{R}^{\alpha(k+m)\dot\alpha(k-m)}
\hat{R}_{\alpha(k+m)\dot\alpha(k-m)}.
\end{equation}
Hence, all the 1-forms
$\Omega^{\alpha(k+m)\dot\alpha(k-m)}$, $|m|\ge n-1$ decouple.

Finally, we derive the unfolded equations for the Skvortsov-Vasiliev
approach. First, we rewrite the gauge sector of unfolded equations via
$\hat{R}^{\alpha(k+m)\dot\alpha(k-m)}$ dropping off all the
decoupled curvatures:
\begin{eqnarray}
\label{SVbeq}
\hat{R}^{\alpha(k+m)\dot\alpha(k-m)} &=& 0, \qquad k<s-1, \nonumber
\\
\hat{R}^{\alpha(s-1+m)\dot\alpha(s-1-m)} &=& 0, \qquad m<n-2, 
\\
\hat{R}^{\alpha(s+n-3)\dot\alpha(s-n+1)} &=& 2E_{\alpha(2)}
W^{\alpha(s+n-1)\dot\alpha(s-n+1)}. \nonumber
\end{eqnarray}
The last equation is the only link between gauge sector and infinite
tail of gauge-invariant zero-forms $W^{\alpha(k+m)\dot\alpha(k-m)}$,
$k\ge s$, $|m| \ge n-1$. The equations for the gauge-invariant forms
remain the same. It is the expected result: since the derivation of
the Skvortsov-Vasiliev-like description is reduced to the fixing the
gauge, the gauge-invariant forms are left unaltered.

\section{Fermionic case}

\subsection{The Lagrangian}

The massless fermion with the spin $s+\frac{1}{2}$ requires only the
physical one-form $ \Psi^{\alpha(s)\dot\alpha(s-1)}$ with its
hermitian conjugate. The Lagrangian for the massless 
spin-$s+\frac{1}{2}$ fermion is:
\begin{eqnarray}
\label{mlfLag}
(-1)^{s}\mathcal{L}^{(s+\frac{1}{2})} &=&
 \Psi_{\alpha(s-1)\beta\dot\alpha(s-1)}
e^{\beta}{}_{\dot\beta} D\Psi^{\alpha(s-1)\dot\alpha(s-1)\dot\beta}
\nonumber
\\
 && + \frac{\lambda(s+1)}{2} \Psi_{\alpha(s-1)\beta\dot\alpha(s-1)}
E^{\beta}{}_{\gamma} \Psi^{\alpha(s-1)\gamma\dot\alpha(s-1)}
\nonumber
\\
 && - \frac{\lambda(s-1)}{2} 
\Psi_{\alpha(s-1)\dot\beta\dot\alpha(s-2)}
E^{\dot\beta}{}_{\dot\gamma}
\Psi^{\alpha(s)\dot\gamma\dot\alpha(s-2)} + h.c.
\end{eqnarray} 
The Lagrangian possesses the following gauge symmetries:
\begin{equation}
\delta \Psi^{\alpha(s){\dot{\alpha}}(s-1)} =
D \eta^{\alpha(s)\dot\alpha(s-1)} + s \lambda 
e^{\alpha}{}_{\dot\alpha} \eta^{\alpha(s-1)\dot\alpha(s)}
+ (s-1) e_{\alpha}{}^{\dot\alpha} \eta^{\alpha(s+1)\dot\alpha(s-2)}.
\end{equation}

The Lagrangian for the massive fermion is built in the same way as for
the boson \cite{Met06,Zin08b}. One introduces the $s+1$ massless
fields for the components with spins $\frac{1}{2}, \frac{3}{2}, \ldots
s+\frac{1}{2}$. The spin-$\frac{1}{2}$ component is described by the
fermionic zero-form $\psi^\alpha$ (with its conjugate), while the
other components require the one-forms 
$\Psi^{\alpha(k)\dot\alpha(k-1)}$ (with their conjugates),
$1 \le k \le s$, used to describe the massless spin-$k+\frac{1}{2}$
fields. The Lagrangian is built as a sum of the massless Lagrangians
with all possible cross-terms and mass-like terms:
\begin{equation}
\label{fLag}
\mathcal{L} = \mathcal{L}_0 + \mathcal{L}_1 
\end{equation}
\begin{eqnarray}
\mathcal{L}_0 &=& \sum_{k=0}^{s-1}
(-1)^{k+1} \Psi_{\alpha(k)\beta\dot\alpha(k)} e^{\beta}{}_{\dot\beta}
D \Psi^{\alpha(k)\dot\alpha(k)\dot\beta} - \alpha_0^2\psi_\alpha
E^{\alpha}{}_{\dot\alpha} D \psi^{\dot\alpha}, 
\\
\mathcal{L}_1 &=& \sum_{k=1}^{s-1} (-1)^{k+1} \alpha_{k}
\Psi_{\alpha(k-1)\beta(2)\dot\alpha(k)} E^{\beta(2)}
\Psi^{\alpha(k-1)\dot\alpha(k)} + \alpha_0^2 \Psi_{\alpha}
E^{\alpha}{}_{\dot\alpha} \psi^{\dot\alpha} + h.c. \nonumber
\\
 && + \sum_{k=1}^{s-1} (-1)^{k+1} \frac{\beta_{k+1}}{2}\big[(k+2)
\Psi_{\alpha(k)\beta\dot\alpha(k)} E^{\beta}{}_{\gamma}
\Psi^{\alpha(k)\gamma\dot\alpha(k)} \nonumber
\\
 && \qquad \qquad \qquad 
- k \Psi_{\alpha(k+1)\dot\beta\dot\alpha(k-1)} 
E^{\dot\beta}{}_{\dot\gamma}
\Psi^{\alpha(k+1)\dot\gamma\dot\alpha(k-1)} \big] \nonumber
\\
 && + \beta_{1} \alpha_0^2E \psi_\alpha \psi^\alpha + h.c.
\end{eqnarray}
The Lagrangian is split into $\mathcal{L}_0$ and $\mathcal{L}_1$
according to the order of the terms in derivatives. The field $\psi$
has a non-canonical normalization. The Lagrangian can be used to
describe the infinite-spin particles as well. In this case one has to
take the limit $s\to+\infty$ \cite{Met17,KhZ17}. We require that the
Lagrangian possesses all the gauge symmetries the massless components
possess. The transformation laws of the fields then have to be
modified with the cross-terms and mass-like terms:
\begin{eqnarray}
\delta \Psi^{\alpha(k+1){\dot{\alpha}}(k)} &=&
D \eta^{\alpha(k+1)\dot\alpha(k)}
+ \alpha_{k+1} e_{\alpha\dot\alpha}
\eta^{\alpha(k+2)\dot\alpha(k+1)}
+ \beta_{k+1}(k+1) e^\alpha{}_{\dot\alpha}
\eta^{\alpha(k)\dot\alpha(k+1)} \nonumber
\\
 && + \frac{(k+1)\alpha_{k}}{(k+2)} e^{\alpha\dot\alpha}
\eta^{\alpha(k)\dot\alpha(k-1)} + k e_\alpha{}^{\dot\alpha}
\eta^{\alpha(k+2)\dot\alpha(k-1)}, 
\\
\delta  \psi^{\alpha} &=& \eta^{\alpha}. \nonumber
\end{eqnarray}
Then, the gauge invariance requirement yields the following relations
for the coefficients $\alpha_k,\beta_k$:
\begin{eqnarray}
k\beta_{k} &=& \beta_{k+1}(k+2), \nonumber
\\
\alpha_{k+1}{}^2 &=& \alpha_k{}^2+\lambda^2(2k+3)-\beta_{k+1}{}^2
(2k+3).
\end{eqnarray}
One can see that the coefficients $\alpha_k,\beta_k$ are defined up to
two free parameters. In the case of the finite spin $s+\frac{1}{2}$ an
additional condition $\alpha_{s}=0$ reduces the number of the free
parameters to one. Similarly to the bosonic case, we choose the "mass"
parameter as this free parameter. It has to be proportional to the
$\alpha_{s-1}$ and tends to the usual mass in the flat-space limit 
$\lambda^2\to 0$. This leads to:
$$
M^2 = \frac{s^2\alpha_{s-1}{}^2}{(2s-1)}.
$$
This gives the following expressions for the coefficients
$\alpha_k,\beta_k$:
\begin{eqnarray}
\alpha_k{}^2 &=& \frac{(s-k)(s+k+2)}{(k+1)^2}
\big[M^2+(s+k+1)(s-k-1)\lambda^2\big], \nonumber
\\
\beta_k{}^2 &=& \frac{(s+1)^2}{k^2(k+1)^2}\big[M^2+s^2\lambda^2\big].
\end{eqnarray}
Such parametrization is not applicable in the case of the infinite
spin. As in the bosonic case, we choose the lower coefficients as the
free parameters - namely, $\beta_1$ and $\alpha_0$:
\begin{eqnarray}
\alpha_k{}^2 &=& \frac{1}{(k+1)^2}
\big[\alpha_0{}^2(k+1)^2-4\beta_1{}^2k(k+2)+k(k+1)^2(k+2)\lambda^2\big], \nonumber
\\
\beta_k{}^2 &=& \frac{4\beta_1{}^2}{k^2(k+1)^2}.
\end{eqnarray}

Let us study the hermiticity of the Lagrangian now; consider the
finite spin $s+\frac{1}{2}$ first. The hermiticity condition is that
all $\alpha_k{}^2$, $\beta_k{}^2$ must be non-negative.
That requires $M^2\ge 0$ in case of flat space and AdS. The case of
$M^2=0$ corresponds to the decoupling of the highest field; moreover,
the Lagrangian breaks component-wise into $s+1$ pieces in the flat
space case. In $dS$ the hermiticity condition leads to the unitary
forbidden region $M^2 < -s^2\lambda^2$. Similarly to the bosonic case,
inside the forbidden region there exists a number of partially
massless limits 
$$
M^2=-\lambda^2 (s+k-1)(s-k+1),
$$
however, none of them are unitary, since the fermionic Lagrangian
contains $\beta_k$ and they are all imaginary in this cases.

Consider the infinite spin case now. We introduce the variable
$y_k=k^2+2k$ for simplicity. Then the sign of the coefficients
$\alpha_k{}^2$ is determined by the square trinomial: 
$$
\alpha_k{}^2 \propto \alpha_0^2+
(\alpha_0{}^2+\lambda^2-4\beta_1^2)y_k+y_k^2\lambda^2.
$$
The sign of $\beta_k{}^2$ is determined by that of $\beta_1{}^2$.
It follows immediately that the infinite-spin representations are
completely impossible in the $dS$, since $\alpha_k{}^2 \propto
y_k{}^2\lambda^2<0$ for sufficiently large values of $k$. Consider the
flat space case. The hermiticity condition reads:
\begin{equation}
\alpha_0{}^2 \ge 4\beta_1{}^2 \ge 0.
\end{equation}
In this, the case $\alpha_0{}^2 = 4\beta_1{}^2$ corresponds to the
massless infinite spin particle which can be obtained from the massive
finite spin one by the limit $s \to \infty$, $M \to 0$, $Ms = const$.
No unitary partially massless limits exist in the flat space. 

In the $AdS$ case, similarly to the bosonic case, the unitarity region
in the $\alpha_0{}^2$, $\beta_1{}^2$ parameter space is an infinite
region with piece-wise linear boundary:
\begin{eqnarray}
2\beta_1{}^2 &\in&
\big[(k-1)k(k+1)(k+2)\lambda^2, k(k+1)(k+2)(k+3)\lambda^2\big], \qquad
k\in\mathbb{N}, \nonumber
\\
\alpha_0{}^2 &>&
k(k+2)\big[\frac{4\beta_1{}^2}{(k+1)^2}-\lambda^2\big].
\end{eqnarray}
Moreover, there exists a number of partially massless limits with the
spectrum  $\overline{k+1/2,+\infty}$ for which solutions can also be
written in a form similar to the massive finite spin one:
\begin{eqnarray}
\beta_k{}^2 &=& \frac{(s+1)^2 \hat{M}^2}{k^2(k+1)^2}, \qquad
\hat{M}^2 < (s+1)^2\lambda^2, \nonumber
\\
\alpha_k{}^2 &=& \frac{(k-s)(k+s+2)}{(k+1)^2} [(k+1)^2\lambda^2 -
\hat{M}^2].
\end{eqnarray}

\subsection{Gauge invariant curvatures}

The construction of the complete set of the gauge invariant objects is
similar to the bosonic case.
One needs the complete set of the one-forms 
$\Omega^{\alpha(k+m+1)\dot\alpha(k-m)}$ and zero-forms 
$W^{\alpha(k+m+1)\dot\alpha(k-m)}$, $k\le s-1$, $m\ge 0$ with their
conjugates (see Figure 3a). 
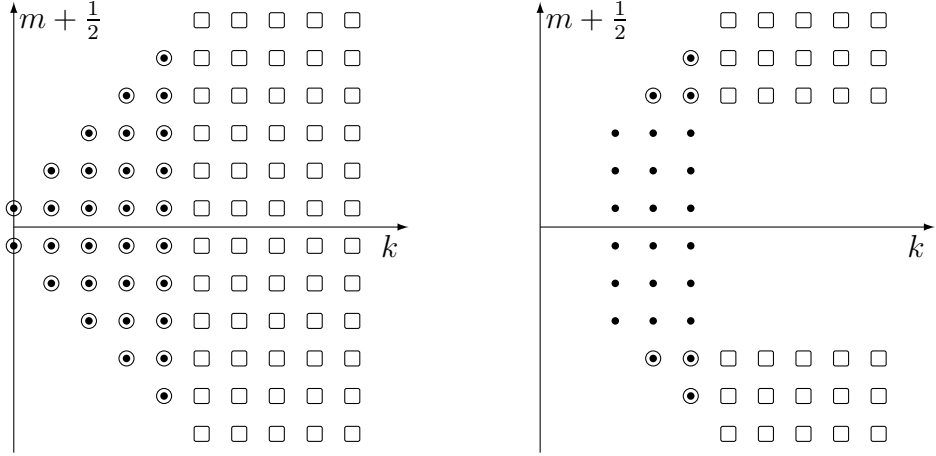
\begin{figure}[htb]
\begin{center}
\begin{picture}(260,120)

\put(10,0){\vector(0,1){120}}
\put(15,110){\makebox(15,10){$m+\frac{1}{2}$}}
\put(10,60){\vector(1,0){105}}
\put(105,50){\makebox(10,10){$k$}}

\multiput(10,55)(0,10){2}{\circle*{2}}
\multiput(10,55)(0,10){2}{\circle{4}}

\multiput(20,45)(0,10){4}{\circle*{2}}
\multiput(20,45)(0,10){4}{\circle{4}}

\multiput(30,35)(0,10){6}{\circle*{2}}
\multiput(30,35)(0,10){6}{\circle{4}}

\multiput(40,25)(0,10){8}{\circle*{2}}
\multiput(40,25)(0,10){8}{\circle{4}}

\multiput(50,15)(0,10){10}{\circle*{2}}
\multiput(50,15)(0,10){10}{\circle{4}}

\multiput(60,115)(10,0){5}{\oval[0.5](4,4)}
\multiput(60,105)(10,0){5}{\oval[0.5](4,4)}
\multiput(60,95)(10,0){5}{\oval[0.5](4,4)}
\multiput(60,85)(10,0){5}{\oval[0.5](4,4)}
\multiput(60,75)(10,0){5}{\oval[0.5](4,4)}
\multiput(60,65)(10,0){5}{\oval[0.5](4,4)}
\multiput(60,55)(10,0){5}{\oval[0.5](4,4)}
\multiput(60,45)(10,0){5}{\oval[0.5](4,4)}
\multiput(60,35)(10,0){5}{\oval[0.5](4,4)}
\multiput(60,25)(10,0){5}{\oval[0.5](4,4)}
\multiput(60,15)(10,0){5}{\oval[0.5](4,4)}
\multiput(60,5)(10,0){5}{\oval[0.5](4,4)}


\put(150,0){\vector(0,1){120}}
\put(155,110){\makebox(15,10){$m+\frac{1}{2}$}}
\put(150,60){\vector(1,0){105}}
\put(245,50){\makebox(10,10){$k$}}

\multiput(170,35)(0,10){6}{\circle*{2}}

\multiput(180,25)(0,10){8}{\circle*{2}}
\put(180,95){\circle{4}}
\put(180,25){\circle{4}}

\multiput(190,15)(0,10){10}{\circle*{2}}
\multiput(190,95)(0,10){2}{\circle{4}}
\multiput(190,15)(0,10){2}{\circle{4}}

\multiput(200,115)(10,0){5}{\oval[0.5](4,4)}
\multiput(200,105)(10,0){5}{\oval[0.5](4,4)}
\multiput(200,95)(10,0){5}{\oval[0.5](4,4)}

\multiput(200,25)(10,0){5}{\oval[0.5](4,4)}
\multiput(200,15)(10,0){5}{\oval[0.5](4,4)}
\multiput(200,5)(10,0){5}{\oval[0.5](4,4)}

\end{picture}
\end{center}
\caption{a) Left figure --- massive fermion with spin 
$s=\frac{11}{2}$. b) Right figure --- partially massless case with
$n=3$.}
\end{figure}
In what follows the notations are unified, i.e. we take 
$\Psi^{\alpha(k+1)\dot\alpha(k)}\equiv
\Omega^{\alpha(k+1)\dot\alpha(k)}$, $\psi^\alpha\equiv W^\alpha$.
The ansatz for the gauge transformations is similar to the bosonic one
due to the multispinor formalism. Namely, the gauge transformations
($m\ge0$) have the form:
\begin{eqnarray}
\delta \Omega^{\alpha(k+m+1)\dot\alpha(k-m)} &=&
D \eta^{\alpha(k+m+1)\dot\alpha(k-m)}
+ (k+m+1)\alpha^{-+}_{k,m} e^\alpha{}_{\dot\alpha}
\eta^{\alpha(k+m+1)\dot\alpha(k-m+1)} \nonumber
\\
 && + (k-m) e_\alpha{}^{\dot\alpha}
\eta^{\alpha(k+m+2)\dot\alpha(k-m-1)}
+ \alpha^{++}_{k,m} e_{\alpha\dot\alpha}
\eta^{\alpha(k+m+2)\dot\alpha(k-m+1)} \nonumber
\\
 && + (k+m+1)(k-m)\alpha^{--}_{k,m} e^{\alpha\dot\alpha}
\eta^{\alpha(k+m)\dot\alpha(k-m-1)}, 
\\
\delta W^{\alpha(k+m+1)\dot\alpha(k-m)} &=&
 \eta^{\alpha(k+m+1)\dot\alpha(k-m)}. \nonumber
\end{eqnarray}
Here
$$
\alpha_{k,0}^{++} = \alpha_{k+1}, \qquad
\alpha_{k,0}^{-+} = \beta_{k+1}, \qquad
\alpha_{k,0}^{--} = \frac{1}{k(k+2)}\alpha_k.
$$
The corresponding expressions for the gauge invariant curvatures are:
\begin{eqnarray}
 R^{\alpha(k+m+1)\dot\alpha(k-m)} &=&
D \Omega^{\alpha(k+m)\dot\alpha(k-m)}
+ (k+m+1)(k-m)\alpha^{--}_{k,m} e^{\alpha\dot\alpha}
\Omega^{\alpha(k+m)\dot\alpha(k-m-1)} \nonumber
\\
 && + \alpha^{++}_{k,m} e_{\alpha\dot\alpha}
\Omega^{\alpha(k+m+2)\dot\alpha(k-m+1)}
+ (k-m) e_\alpha{}^{\dot\alpha}
\Omega^{\alpha(k+m+2)\dot\alpha(k-m-1)} \nonumber
\\
 && + (k+m+1)\alpha^{-+}_{k,m} e^\alpha{}_{\dot\alpha}
\Omega^{\alpha(k+m-2)\dot\alpha(k-m+1)}, \nonumber
\\
 R^{\alpha(2k+1)} &=& D \Omega^{\alpha(2k+1)}
+ \alpha^{++}_{k,k} e_{\alpha\dot\alpha}
\Omega^{\alpha(2k+2)\dot\alpha}
+ (2k+1)\alpha^{-+}_{k,k} e^\alpha{}_{\dot\alpha}
\Omega^{\alpha(2k)\dot\alpha} \nonumber
\\
 && - 4k(2k+1)\alpha^{-+}_{k,k}\alpha^{--}_{k,k-1} E^{\alpha(2)}
W^{\alpha(2k-1)} - 2\alpha^{++}_{k,k} E_{\alpha(2)} W^{\alpha(2k+3)}
\\
 && - \frac{2\alpha^{-+}_{k+1}}{(2k+3)} E^{\alpha}{}_{\beta}
W^{\alpha(2k)\beta}, \nonumber
\\
 C^{\alpha(k+m+1)\dot\alpha(k-m)} &=&
D W^{\alpha(k+m+1)\dot\alpha(k-m)}
+ (k+m+1)(k-m)\alpha^{--}_{k,m} e^{\alpha\dot\alpha}
W^{\alpha(k+m)\dot\alpha(k-m-1)} \nonumber
\\
 && + \alpha^{++}_{k,m} e_{\alpha\dot\alpha}
W^{\alpha(k+m+2)\dot\alpha(k-m+1)}
+ (k-m) e_\alpha{}^{\dot\alpha}
W^{\alpha(k+m+2)\dot\alpha(k-m-1)} \nonumber
\\
 && + (k+m+1)\alpha^{-+}_{k,m} e^\alpha{}_{\dot\alpha}
W^{\alpha(k+m)\dot\alpha(k-m+1)} -
\Omega^{\alpha(k+m+1)\dot\alpha(k-m)}. \nonumber
\end{eqnarray}
It is convenient to express the coefficients $\alpha^{ij}_{k,m}$ via
$\alpha^{++}_{k},\alpha^{--}_{k},\alpha^{-+}_{m}$ (see Appendix B
about the coefficients $\alpha^{ij}_{k,m}$):
\begin{eqnarray}
\alpha^{++}_{k,m} &=& \frac{\alpha^{++}_{k}}{(k-m+1)(k-m+2)},
\nonumber\\
\alpha^{--}_{k,m} &=& \frac{\alpha^{--}_{k}}{(k+m+1)(k+m+2)}, 
\\
\alpha^{-+}_{k,m} &=& \frac{\alpha^{-+}_{m}}
{(k-m+1)(k-m+2)(k+m+1)(k+m+2)} \nonumber
\\
\alpha^{-+}_{k,m} &=& \frac{\alpha^{-+}_{0}}{(k+1)(k+2)}. \nonumber
\end{eqnarray}
We use the same parameter choice for these coefficients, as for the
$\alpha_k$, $\beta_k$. The values of the 
$\alpha^{++}_{k},\alpha^{--}_{k},\alpha^{-+}_{m}$ in case of finite
spin are:
\begin{eqnarray}
\alpha^{++}_{k-1}{}^2 &=& k^2(s-k)(s+k+2)
\big[M^2 + (s+k+1)(s-k-1)\lambda^2\big], \nonumber
\\
\alpha^{--}_{k}{}^2 &=& \frac{1}{k^2}(s-k)(s+k+2)
\big[M^2 + (s+k+1)(s-k-1)\lambda^2\big], \nonumber
\\
\alpha^{-+}_{m} &=& (s-m+1)(s+m+1)\big[M^2 + (s-m)(s+m)\lambda^2\big],
\\
\alpha^{-+}_{0}{}^2 &=& (s+1)^2\big[M^2 + s^2\lambda^2\big]. \nonumber
\end{eqnarray}
In the infinite spin case, their values are:
\begin{eqnarray}
\alpha^{++}_{k-1}{}^2 &=& k^2\big[\alpha_0{}^2(k+1)^2 
- 4\beta_1{}^2k(k+2) + k(k+1)^2(k+2)\lambda^2\big], \nonumber
\\
\alpha^{--}_{k}{}^2 &=& \frac{1}{k^2}
\big[\alpha_0{}^2(k+1)^2 - 4\beta_1{}^2k(k+2) +
k(k+1)^2(k+2)\lambda^2\big], \nonumber
\\
\alpha^{-+}_{k+1} &=& \big[\alpha_0{}^2(k+1)^2 - 4\beta_1{}^2k(k+2)+
k(k+1)^2(k+2)\lambda^2\big], 
\\
\alpha^{-+}_{0}{}^2 &=& 4\beta_1{}^2. \nonumber
\end{eqnarray}
Let us outline the useful relation
$\alpha^{--}_{m-1}\alpha^{++}_{m-2}=\alpha^{-+}_{m}$ once again.

The hermiticity of the curvatures is given by the same expressions as
for the Lagrangian. We have already seen that there the partially
massless limits are non unitary due to the presence of $\beta_k$ in
the finite spin case. In curvatures, the coefficient 
$\alpha^{-+}_{k,0}$ plays the role of $\beta_k$. 

In case of the partially massless limit with the components 
$\overline{k+1/2,s+1/2}$ all the lower spin fields (i.e.
$\Omega^{\alpha(l+m+1)\dot\alpha(l-m)}$,
$W^{\alpha(l+m+1)\dot\alpha(l-m)}$ for $l < k-1$) completely decouple.
Besides, all the zero-forms $W^{\alpha(l+m+1)\dot\alpha(l-m)}$
with $l \ge k-1$, $m \le k-1$ and their conjugates also decouple. This
leaves us with the set of one-forms 
$\Omega^{\alpha(l+m+1)\dot\alpha(l-m)}$ with $l \ge k-1$, $m \le k-1$
and their conjugates (which form an analogue of the Skvortsov-Vasiliev
formalism for the fermions, see below) as well as the pairs of
one-forms and zero-forms with $l \ge k$, $l \ge m \ge k$ (see Figure
3b).

\subsection{Lagrangian in terms of curvatures}

The ansatz for the Lagrangian expressed in the terms of the curvatures
is similar to the bosonic case:
\begin{eqnarray}
\label{fcLag}
\mathcal{L} &=&
\sum_{k=0}^{s-1}\sum_{m=-k-1}^k (-1)^{k+1} a_{k,m}
 R^{\alpha(k+m+1)\dot\alpha(k-m)} R_{\alpha(k+m+1)\dot\alpha(k-m)}
\nonumber
\\
 && + \sum_{k=0}^{s-2}\sum_{m=-k-1}^k (-1)^{k+1}b_{k,m}
 R^{\alpha(k+m+1)\dot\alpha(k-m)} e^{\alpha\dot\alpha}
C_{\alpha(k+m+2)\dot\alpha(k-m+1)} \nonumber
\\
 && + \sum_{k=0}^{s-1}\sum_{m=-k}^{k-1} (-1)^{k+1}c_{k,m}
 R^{\alpha(k+m+1)\dot\alpha(k-m)} e_{\alpha\dot\alpha}
C_{\alpha(k+m)\dot\alpha(k-m-1)} \nonumber
\\
 && + \sum_{k=0}^{s-1}\sum_{m=-k-1}^{k} (-1)^{k+1}d_{k,m}
 R^{\alpha(k+m+1)\dot\alpha(k-m)} e^\beta{}_{\dot\alpha}
C_{\alpha(k+m+1)\beta\dot\alpha(k-m)} \nonumber
\\
 && + \sum_{k=0}^{s-1}\sum_{m=-k}^{k+1} (-1)^{k+1}d_{k,-m}
 R^{\alpha(k+m)\dot\alpha(k-m+1)} e_\alpha{}^{\dot\beta}
C_{\alpha(k+m)\dot\beta\dot\alpha(k-m+1)} \nonumber
\\
 && + \sum_{k=0}^{s-1}\sum_{m=-k}^k (-1)^{k+1}e_{k,m}
 C^{\alpha(k+m+1)\dot\alpha(k-m)} E^\beta{}_\alpha
C_{\alpha(k+m)\beta\dot\alpha(k-m)} \nonumber
\\
 && + \sum_{k=0}^{s-1}\sum_{m=-k}^{k} (-1)^{k+1}e_{k,-m}
 C^{\alpha(k+m)\dot\alpha(k-m+1)} E^{\dot\beta}{}_{\dot\alpha}
C_{\alpha(k+m)\dot\beta\dot\alpha(k-m)} \nonumber
\\
 && + \sum_{k=0}^{s-2}\sum_{m=-k-1}^k (-1)^{k+1}f_{k,m}
 C^{\alpha(k+m+1)\dot\alpha(k-m)} E^{\alpha(2)}
C_{\alpha(k+m+3)\dot\alpha(k-m)} \nonumber
\\
 && + \sum_{k=0}^{s-2}\sum_{m=-k-1}^k (-1)^{k+1}f_{k,-m}
 C^{\alpha(k+m)\dot\alpha(k-m+1)} E^{\dot\alpha(2)}
C_{\alpha(k+m)\dot\alpha(k-m+3)}.
\end{eqnarray}
Here $a_{k,m}=a_{k,-m-1}$, $b_{k,m}=b_{k,-m-1}$, $c_{k,m}=c_{k,-m-1}$
due to the hermiticity. 

We determine the coefficients $ a_{k,m}-f_{k,m}$ in the same way as
for the boson, i.e. we require that the extra fields decouple and the
equations of motion derived from (\ref{fcLag}) and (\ref{fLag})
match. The equation of motion obtained from (\ref{fLag}) are expressed
via curvatures as:
\begin{eqnarray}
\frac{\delta\mathcal{L}}{\delta \Omega^{\alpha(k+1)\dot\alpha(k)}}
&=& (-1)^{k+1} e_\alpha{}^{\dot\alpha} 
R_{\alpha(k)\dot\alpha(k+1)}, \nonumber
\\
\frac{\delta\mathcal{L}}{\delta  \Omega^{\alpha}} &=&
 e_\alpha{}^{\dot\alpha} R_{\dot\alpha} 
\\
\frac{\delta\mathcal{L}}{\delta W^\alpha} &=&
- \alpha_0{}^2 E_{\alpha}{}^{{\dot{\alpha}}} C_{\dot\alpha}. \nonumber
\end{eqnarray}
As in the bosonic case, there is an arbitrarity in the choice for the
coefficients $a_{k,m}-f_{k,m}$. It comes from the fact that the
Lagrangian (\ref{fcLag}) is defined up to the total derivatives:
\begin{eqnarray}
(\mathcal{L}-\mathcal{L}_0) &=&
\sum_{k=0}^{s-1}\sum_{m=0}^k (-1)^{k+1} p_{k,m} D(
R^{\alpha(k+m+1)\dot\alpha(k-m)}
C_{\alpha(k+m+1)\dot\alpha(k-m)}+h.c.) \nonumber
\\
 && + \sum_{k=0}^{s-1}\sum_{m=0}^{k} (-1)^{k+1} q_{k,m} D(
C^{\alpha(k+m+1)\dot\alpha(k-m)} e^\alpha{}_{\dot\alpha}
C_{\alpha(k+m+2)\dot\alpha(k-m)}+h.c.) \nonumber
\\
 && + \sum_{k=0}^{s-1}\sum_{m=0}^k (-1)^{k+1} r_{k,m} D(
C^{\alpha(k+m+1)\dot\alpha(k-m)} e^{\alpha\dot\alpha}
C_{\alpha(k+m+2)\dot\alpha(k-m+1)} + h.c.)
\end{eqnarray}
Those terms, however, lead to the shifts of the coefficients 
$a_{k,m}-f_{k,m}$, just like in the bosonic case (see
Appendix B for the explicit expressions of these shifts). 
Similarly to the bosonic case, we choose the coefficients
$p_{k,m}$, $q_{k,m}$ and $r_{k,m}$ so that all the $b_{k,m}$, 
$c_{k,m}$, $d_{k,m}$ and $e_{k,m}$, $f_{k,m}$ for $k\ne m$ equal to
zero. The expressions for the nonzero coefficients read:
\begin{eqnarray}
\label{fsps}
a_{k,m}^{(0)} &=&
\frac{(k+m+2)!k!^2}{4(k-m)!^2(k-m+1)!\prod_{i=0}^{m}\alpha^{-+}_{i}},
\nonumber
\\
e_{k,k}^{(0)} &=&
\frac{(2k+2)!k!^2\alpha^{-+}_{k+1}}{2(2k+3)\prod_{i=0}^{m}
\alpha^{-+}_{i}}, 
\\ 
f_{k,k}^{(0)} &=& -
\frac{\alpha^{++}_{k}(2k)!k!^2}{\prod_{i=0}^{m}\alpha^{-+}_{i}}.
\nonumber
\end{eqnarray}
A general solution $a_{k,m}-f_{k,m}$ can be obtained from
the special solution (\ref{fsps}) by substituting the arbitrary
$p_{k,m}$, $q_{k,m}$, $r_{k,m}$ in (\ref{genfs}).

We use the same procedure to eliminate the singularities in the
coefficients $a_{k,m}-f_{k,m}$ in the case of partially massless limit
given by $\alpha^{-+}_{n}=\alpha^{--}_{n-1}=\alpha^{++}_{n-2}=0$. Let
us recall the steps. First, we obtain the general solution
$a_{k,m}-f_{k,m}$ for arbitrary parameters $s,M$
($\alpha_0,\beta_1$). Then we choose $p_{k,m}$, $q_{k,m}$, $r_{k,m}$
in a way that allows to take the limit $\alpha^{-+}_{n}=0$. The
easiest way is to zero out all the "bad" coefficients preserving most
$b_{k,m}$, $c_{k,m}$, $d_{k,m}$ and $e_{k,m}$, $f_{k,m}$ zero. The
exact expressions for the non-zero $p_{k,m}$, $q_{k,m}$ are (all
$r_{k,m}$ are zero) are given in Appendix B. The expressions for the
coefficients with $k<n$ remain the same, except the vanishing
coefficients $e_{n-1,n-1}$, $f_{n-2,n-2}$, $f_{n-1,n-1}$. The
expressions for the non-zero coefficients for  $k\ge n$ are:
\begin{eqnarray}
a_{k,\pm m} &=&
\frac{(k+m+2)!k!^2}{4(k-m)!^2(k-m+1)!\prod_{i=0}^{m}\alpha^{-+}_{i}},
\qquad m<n, \nonumber
\\
d_{k,n-1} &=& - \frac{(k+n+1)!k!^2}{2(k-n)!(k-n+1)!(k-n+2)!
\prod_{i=0}^{n-1}\alpha^{-+}_{i}}, \nonumber
\\
e_{k,n} &=&
-\frac{(k+n+1)!k!^2}{4(k-n)!(k-n+1)!^2\prod_{i=0}^{n-1}
\alpha^{-+}_{i}}, 
\\
e_{k,-n-1} &=& 
\frac{(k+n+1)!k!^2}{4(k-n-1)!(k-n+1)!(k-n+2)!\prod_{i=0}^{n-1}\alpha^{-+}_{i}}. \nonumber
\end{eqnarray}
The Lagrangian is thus split into two parts, one containing the
coefficients with $k\ge n-1$ and the other containing the ones with
$k<n-1$:
\begin{eqnarray}
-i\mathcal{L} &=& -i\mathcal{L}^{(0,n-2)} \nonumber
\\
 && + \sum_{k=n-1}^{s-1}\sum_{m=-n}^{n-1} (-1)^{k+1} a_{k,m}
 R^{\alpha(k+m+1)\dot\alpha(k-m)} R_{\alpha(k+m+1)\dot\alpha(k-m)}
\nonumber
\\
 && + \sum_{k=n}^{s-1} (-1)^{k+1}d_{k,n-1}\big[
 R^{\alpha(k+n)\dot\alpha(k-n+1)} e^\beta{}_{\dot\alpha}
C_{\alpha(k+n)\beta\dot\alpha(k-n)} - h.c. \big] \nonumber
\\
 && + \sum_{k=n}^{s-1} (-1)^{k+1}e_{k,n} \big[
 C^{\alpha(k+n+1)\dot\alpha(k-n)} E^\beta{}_\alpha
C_{\alpha(k+n)\beta\dot\alpha(k-n)} - h.c. \big] \nonumber
\\
 && + \sum\limits_{k=n}^{s-1} (-1)^{k+1}e_{k,-n-1}
\big[ C^{\alpha(k+n+1)\dot\alpha(k-n)}
E^{\dot\beta}{}_{\dot\alpha}
C_{\alpha(k+n+1)\dot\beta\dot\alpha(k-n-1)} - h.c. \big].
\end{eqnarray}
One can see that the part with the higher components does not contain
the curvatures which contain the fields $
W^{\alpha(k+m+1){\dot{\alpha}}(k-m)}$, $m \le n-1$. That means that
the part with the higher coefficients does not contain the components
$\overline{\frac{1}{2},n-\frac{1}{2}}$, i.e. the components
$\overline{n+\frac{1}{2},s+\frac{1}{2}}$ decouple in this case as
well. In case of $n=0$ the expression is:
\begin{eqnarray}
-i\mathcal{L} &=& \sum_{k=0}^{s-1} (-1)^{k+1}d_{k,-1}\big[
 R^{\alpha(k)\dot\alpha(k+1)} e^\beta{}_{\dot\alpha}
C_{\alpha(k)\beta\dot\alpha(k)} - h.c. \big] \nonumber
\\
 && + \sum\limits_{k=0}^{s-1} (-1)^{k+1}e_{k,n} \big[
 C^{\alpha(k+n+1)\dot\alpha(k-n)} E^\beta{}_\alpha
C_{\alpha(k)\beta\dot\alpha(k)} - h.c. \big] \nonumber
\\
 && + \sum_{k=0}^{s-1} (-1)^{k+1}e_{k,-n-1} \big[
 C^{\alpha(k+1)\dot\alpha(k)} E^{\dot\beta}{}_{\dot\alpha}
C_{\alpha(k+1)\dot\beta\dot\alpha(k-1)} - h.c. \big]
\end{eqnarray}
This case correspond to unitary region boundary, when the curvatures
contain the fields $ \Omega^{\alpha(k+m+1)\dot\alpha(k-m)},
W^{\alpha(k+m+1)\dot\alpha(k-m)}$ with $m\ge 0$ or with $m<0$
only. It does not correspond to any partially massless limit.

\subsection{Unfolded equations}

We derive the unfolded equations chain in the same way as in the
bosonic case. We start by setting to zero most of the gauge invariant
curvatures:
\begin{eqnarray}
\label{fUnf}
0 &=& R^{\alpha(s+m)\dot\alpha(s-m-1)}, \qquad m \ne s-1, -s,
\nonumber
\\
0 &=& R^{\alpha(k+m+1)\dot\alpha(k-m)}, \qquad k < s-1,,
\\
0 &=& C^{\alpha(k+m+1)\dot\alpha(k-m)}, \qquad k < s-1. \nonumber
\end{eqnarray}
To construct the consistent equations for the remaining gauge
invariant curvatures $ R^{\alpha(2s-1)}$ and
$C^{\alpha(s+m)\dot\alpha(s-m-1)}$ one has to introduce a first set of
the gauge invariant zero-forms:
\begin{eqnarray}
\label{fUnf1}
0 &=& R^{\alpha(2s-1)} - 2 E_{\alpha(2)} W^{\alpha(2s+1)}, \nonumber
\\
0 &=& C^{\alpha(s+m)\dot\alpha(s-m-1)} + e_{\alpha\dot\alpha}
W^{\alpha(s+m+1)\dot\alpha(s-m)}. 
\end{eqnarray}
The first equation starts the zero-form chain for the highest-spin
massless component while the other extend the zero-form chains for the
components with spins $\overline{\frac{1}{2},s-\frac{1}{2}}$ (see
Figure 3a). The most general ansatz (up to the normalization) for the
infinite tail containing the gauge-invariant zero-forms is:
\begin{eqnarray}
\label{fUnf2}
0 &=& D W^{\alpha(k+m+1)\dot\alpha(k-m)} + (k+m+1)(k-m)
\beta^{--}_{k,m} e^{\alpha\dot\alpha} W^{\alpha(k+m)\dot\alpha(k-m-1)}
\nonumber
\\
 && + e_{\beta\dot\beta} 
 W^{\alpha(k+m+1)\beta\dot\alpha(k-m)\dot\beta} 
 + (k-m)\beta^{+-}_{k,m} e_\beta{}^{\dot\alpha}
W^{\alpha(k+m+1)\beta\dot\alpha(k-m-1)} \nonumber
\\
 && + (k+m+1)\beta^{-+}_{k,m} e^\alpha{}_{\dot\beta}
W^{\alpha(k+m)\dot\alpha(k-m)\dot\beta}.
\end{eqnarray}
Here $\beta^{-+}_{k,m}=\beta^{+-}_{k,-m-1}$ due to hermiticity. The
equations (\ref{fUnf1}) and (\ref{fUnf2}) must agree with each other;
the consistency requirement yields the following expression for the
$\beta^{ij}_{k,m}$:
\begin{eqnarray}
\beta^{-+}_{k,m} &=& \frac{\beta^{-+}_{m}}{(k+m+2)(k+m+1)}, \nonumber
\\
\beta^{+-}_{k,m} &=& \frac{\beta^{+-}_{m}}{(k-m+1)(k-m)}, \nonumber
\\
\beta^{--}_{k,m} &=& \frac{\alpha^{-+}_{k+1}}
{(k+m+1)(k+m+2)(k-m)(k-m+1)}, \nonumber
\\
\beta^{-+}_{m} &=& \frac{\alpha^{-+}_{m}}{(s-m)(s-m+1)}, \qquad 
1\le m<s, \nonumber
\\
\beta_{s}^{-+} &=& \frac{\alpha^{-+}_s}{2}, \qquad
\beta^{-+}_{0} = \alpha^{-+}_{0}, \nonumber
\\
\beta^{+-}_{m} &=& (s-m-1)(s-m), \quad 1\le m<s-1, \qquad
\beta^{+-}_{s-1} = 2.
\end{eqnarray}

As in the bosonic case in the partially massless limit
$\alpha^{-+}_{n}=0$ curvatures $
C^{\alpha(k+m+1)\dot\alpha(k-m)}$, $m<n$ and fields 
$W^{\alpha(k+m+1)\dot\alpha(k-m)}$, $m<n$ decouple, as a result
all equations, containing 
$W^{\alpha(k+m+1)\dot\alpha(k-m)}$, $m<n$ completely decouple (see
Figure 3b).

\subsection{Skvortsov-Vasiliev formalism for fermions}

The Skvortsov-Vasiliev formalism \cite{SV06} can be extended to the
fermionic partially massless particles. Consider the
$\overline{n+1/2,s+1/2}$-partially massless limit. We start with the
partial fixing of the gauge by setting 
$W^{\alpha(k+m+1)\dot\alpha(k-m)}=0$, $m< n-1$. Then, we introduce
the modified 2-curvature 
$\hat{R}^{\alpha(k+n-1)\dot\alpha(k-n+2)}$ as:
\begin{eqnarray}
 \hat{R}^{\alpha(k+n-1)\dot\alpha(k-n+2)} &=&
R^{\alpha(k+n-1)\dot\alpha(k-n+2)} + (k-n+2)
e_\alpha{}^{\dot\alpha} C^{\alpha(k+n)\dot\alpha(k-n+1)} \nonumber 
\\
 &=& D \Omega^{\alpha(k+n-1)\dot\alpha(k-n+2)}
+ \alpha^{++}_{k,n-2} e_{\alpha\dot\alpha}
\Omega^{\alpha(k+n)\dot\alpha(k-n+3)} \nonumber
\\
 && + (k+n-1)(k-n+2)\alpha^{--}_{k,n-2} e^{\alpha\dot\alpha}
\Omega^{\alpha(k+n)\dot\alpha(k-n+1)} \nonumber
\\
 && + (k+n-1)\alpha^{-+}_{k,n-2} e^\alpha{}_{\dot\alpha}
\Omega^{\alpha(k+n-2)\dot\alpha(k-n+3)}.
\end{eqnarray}
For simplicity we unify the notation defining 
$\hat{R}^{\alpha(k+m+1)\dot\alpha(k-m)}=
R^{\alpha(k+m+1)\dot\alpha(k-m)}$, $m<n-2$. Then, we reformulate
the theory in terms of the modified curvatures 
$\hat{R}^{\alpha(k+m+1)\dot\alpha(k-m)}$, i.e. without zero-forms
and the 1-forms $ \Omega^{\alpha(k+m+1)\dot\alpha(k-m)}$, 
$|m|\ge n-1$. Then we can express the Lagrangian via the modified
curvatures as follows:
\begin{equation}
-i\mathcal{L} = -i\mathcal{L}^{(0,n-2)} +
\sum_{k=n-1}^{s}\sum_{m=-n+1}^{n-1} (-1)^{k+1} a_{k,m}
 \hat{R}^{\alpha(k+m+1)\dot\alpha(k-m)}
\hat{R}_{\alpha(k+m+1)\dot\alpha(k-m)}.
\end{equation}

The derivation of the unfolded equations is straightforward. For this,
we rewrite the gauge sector of unfolded equations via $\hat{R}$
dropping all the decouples curvatures:
\begin{eqnarray}
\hat{R}^{\alpha(k+m+1)\dot\alpha(k-m)} &=& 0, \qquad k < s-1,
\nonumber
\\
\hat{R}^{\alpha(s+m)\dot\alpha(s-1-m)} &=& 0, \qquad m < n-2,
\\
\hat{R}^{\alpha(s+n-2)\dot\alpha(s-n+1)} &=& 2E_{\alpha(2)}
W^{\alpha(s+n)\dot\alpha(s-n+1)}. \nonumber
\end{eqnarray}
The last equation is the only link between the gauge sector and the
sector of the gauge invariant zero-forms
$W^{\alpha(k+m+1)\dot\alpha(k-m)}$, $k > s$, $m \ge n-1$ and their
conjugates. In this, the equations for these gauge invariant
zero-forms remains to be the same.

\section{Conclusion}

In the paper, the gauge invariant description of the massive higher
spin bosons and fermions was built in (A)dS${}_4$. In both cases, we
begin with the construction of the gauge invariant Lagrangian and
investigate an unitarity of models obtained, including all possible
partially massless and/or infinite spin limits. For both bosons and
fermions, a complete set of the gauge invariant curvatures was
constructed (introducing all necessary extra fields) and the
Lagrangian was expressed via these curvatures. At last, the complete
set of the unfolded equations was constructed. Also, the connection
with the Skvortsov-Vasiliev formalism \cite{SV06} was
discussed; it was shown that such formalism can be obtained by the
partial gauge fixing to get rid of the zero-forms that do not decouple
in the partially massless limit. As a byproduct, we obtain  the
unfolded equations for the Skvortsov-Vasiliev formalism. Moreover,
we have shown that the analogous formalism exists for the partially
massless fermions as well. The calculations were carried out in the
multispinor formalism, which enabled us to simplify the formulae at
the cost of the restriction to $d=4$. Multispinor formalism also
treats the bosons and fermions in a similar way, which make the work
particularly useful for the supersymmetry studies. The results were
already used in the studies of the different higher spin $N=1$
supermultiplets in $d=4$ \cite{BKhSZ19,BKhSZ19a,BKhSZ19b}.

\appendix

\section{Notations  and conventions}

We work in the frame-like multispinor formalism. It means that all
objects are forms with multispinors as their local indices, i.e.
$\Phi^{\alpha(k)\dot\alpha(k)}$. World indices are omitted everywhere;
all expressions are completely antisymmetric on them. We use the
condensed index notations. Namely, if the expression is symmetric on
upper/low indices $\alpha_1\alpha_2\cdots\alpha_k$, these indices are
denoted with the same letter with the number of indices in
parentheses. For example:
$$
R^{\alpha_1\alpha_2\cdots\alpha_s}= R^{\mu(s)}.
$$ 
Symmetrization over the set of $n$ indices is defined as the sum of
all the $n!$ expressions obtained from the initial one by all the
possible permutations of these indices, with the normalization factor
$1/n!$. According to the definition of the symmetrization, multiple
symmetrization over the same set of indices is equivalent to the
unique symmetrization. For example: 
\begin{eqnarray*}
 R^{\alpha(s)} R_{\alpha(s)} a^\alpha a_\alpha &=& \frac{s-1}{s}
R^{\alpha(s-1)\beta} R_{\alpha(s)} a^\alpha a_\beta + \frac{1}{s}
R^{\alpha(s)} R_{\alpha(s)} a^\beta a_\beta, 
\\
 R^{\alpha(s-1)\beta} R_{\alpha(s)} a^\alpha a_\beta &=& 
R^{\alpha(s-1)\beta} R_{\alpha(s-1)\gamma} a^\gamma a_\beta.
\end{eqnarray*} 
The indices are contracted according to the Einstein rule, with the
respect to the symmetrization. For example:
$$
 e_\alpha A^\alpha B^{\alpha(2)} \equiv
\frac{1}{3} (e_{\beta} A^{\beta} B^{\alpha(2)}+2 e_{\beta} A^{\alpha}
B^{\alpha\beta}).
$$ 
In four-dimensional space-time, using the Lorenz algebra isomorphism
$\mathfrak{so}(3,1)\sim\mathfrak{sl}(2,\mathbb{C})$ one can replace
vector index by two spinor indices with values $i=1,2$ \cite{DS14}:
$T^\mu\sim  T^{\alpha\dot\alpha}$. The spinor indices are raised and
lowered with the antisyymetric tensors $ \epsilon_{\alpha\beta}$ 
($\epsilon_{\dot\alpha\dot\beta}$):
\begin{equation}
 \epsilon_{\alpha\beta} \xi^\beta = - \xi_\alpha, \qquad
 \epsilon^{\alpha\beta} \xi_\beta = \xi^\alpha,
\end{equation} 
the same is true for dotted indices. Hence, all the symmetric
multispinors are automatically traceless. Under the Hermitian
conjugation, dotted and undotted indices are transformed one into
another. For example:
$$
\left( A^{\alpha(2)\dot\beta}\right)^\dagger = A^{\beta\dot\alpha(2)}.
$$ 
The mixed symmetry tensor $ \Phi^{\mu(k),\nu(l)}$ which corresponds to
the two-row Young tableaux $Y(k,l)$ \cite{BB06} in multispinor
formalism is described by a pair of multispinors 
$\Phi^{\alpha(k+l)\dot\alpha(k-l)}$,
$\Phi^{\alpha(k-l)\dot\alpha(k+l)}$. If the tensor 
$\Phi^{\mu(k),\nu(l)}$ is real then:
\begin{equation}
\left(\Phi^{\alpha(k+l)\dot\alpha(k-l)}\right)^\dagger =
\Phi^{\alpha(k-l)\dot\alpha(k+l)}.
\end{equation} 
Similarly, the mixed symmetry spin-tensor $ \Psi^{\mu(k),\nu(l)}$
which corresponds to the Young tableaux $Y(k+1/2,l+1/2)$ is described
by a pair of multispinors 
$\Psi^{\alpha(k+l+1)\dot\alpha(k-l)}$,
$\Psi^{\alpha(k-l+1)\dot\alpha(k+l)}$. If the spin-tensor 
$\Psi^{\mu(k),\nu(l)}$ is Majorana one then
\begin{equation}
\left(\Psi^{\alpha(k+l+1)\dot\alpha(k-l)}\right)^\dagger =
\Psi^{\alpha(k-l)\dot\alpha(k+l+1)}.
\end{equation} 
The fermionic fields are grassmanian, i.e. they anticommute.

The $AdS_4$ space is described by the background Lorentz connections
$\omega^{\alpha(2)}$, $\omega^{\dot\alpha(2)}$, which enter
implicitly through the Lorentz covariant derivative $D$, and the
background frame $e^{\alpha\dot\alpha}$. We also use the basis
elements for the two-, thee- and four-forms
\begin{equation}
 e^a \sim e^{\alpha\dot\alpha}, \qquad 
 E^{ab} \sim E^{\alpha(2)}, E^{\dot\alpha(2)}, \qquad
 E^{abc} \sim E^{\alpha\dot\alpha}, \qquad 
 E^{abcd} \sim E,
\end{equation} 
defined as follows:
\begin{eqnarray}
e^{\alpha\dot\alpha} \wedge e^{\beta\dot\beta} &=&
\varepsilon^{\alpha\beta} E^{\dot\alpha\dot\beta}
+ \varepsilon^{\dot\alpha\dot\beta} E^{\alpha\beta}, \nonumber
\\
E^{\alpha(2)} \wedge e^{\beta\dot\alpha} &=& \varepsilon^{\alpha\beta}
E^{\alpha\dot\alpha}, \\
E^{\alpha\dot\alpha} \wedge e^{\beta\dot\beta} &=&
\varepsilon^{\alpha\beta} \varepsilon^{\dot\alpha\dot\beta} E.
\nonumber
\end{eqnarray}
The hermitian conjugation rules for the basis forms are:
\begin{equation}
\left( e^{\alpha\dot\alpha}\right)^\dagger = e^{\alpha\dot\alpha},
\qquad \left( E^{\alpha(2)}\right)^\dagger = E^{\dot\alpha(2)}, \qquad
\left( E^{\alpha\dot\alpha}\right)^\dagger = - E^{\alpha\dot\alpha},
\qquad \left(E\right)^\dagger = - E.
\end{equation} 
The Lorentz covariant derivative is normalized so that
\begin{equation}
D \wedge D \Phi^{\alpha(k)\dot\alpha(l)} = - 2\lambda^2 [(k+m)
E^\alpha{}_\beta \Phi^{\alpha(k-1)\beta} + (k-m)
E^{\dot\alpha}{}_{\dot\beta} 
\Phi^{\alpha(k)\dot\alpha(l-1)\dot\beta}].
\end{equation}
The parameter $\lambda^2$ is proportional to the curvature of the
space-time. The $AdS$ space has $\lambda^2>0$, while the $d$S space
has
$\lambda^2<0$. The case of $\lambda^2=0$ corresponds to the flat
Minkowski space.

In the main text all the wedge product signs $\wedge$ are omitted.

\section{Relations for gauge invariant curvatures}

Firstly, consider the following problem. A set of objects $
B^{\alpha(k+m){\dot{\alpha}}(k-m)}$ is given. Each object has the
form:
\begin{eqnarray}
 B^{\alpha(k+m)\dot\alpha(k-m)} &=& D
A^{\alpha(k+m)\dot\alpha(k-m)}
+ (k+m)(k-m)\alpha^{++}_{k,m} e^{\alpha\dot\alpha}
A^{\alpha(k+m-1)\dot\alpha(k-m-1)} \nonumber
\\
 && + \alpha^{--}_{k,m} e_{\alpha\dot\alpha}
A^{\alpha(k+m+1)\dot\alpha(k-m+1)}
+ (k+m)\alpha^{-+}_{k,m} e^\alpha{}_{\dot\alpha}
A^{\alpha(k+m-1)\dot\alpha(k-m+1)} \nonumber
\\
 && + (k-m)\alpha^{+-}_{k,m} e_\alpha{}^{\dot\alpha}
A^{\alpha(k+m+1)\dot\alpha(k-m-1)},
\end{eqnarray} 
and for each $k,m$ the following relation holds:
\begin{eqnarray}
0 &=& D B^{\alpha(k+m)\dot\alpha(k-m)}
+ (k+m)(k-m)\beta^{++}_{k,m} e^{\alpha\dot\alpha}
B^{\alpha(k+m-1)\dot\alpha(k-m-1)} \nonumber
\\
 && + \beta^{--}_{k,m} e_{\alpha\dot\alpha}
B^{\alpha(k+m+1)\dot\alpha(k-m+1)}
+ (k+m)\beta^{-+}_{k,m} e^\alpha{}_{\dot\alpha}
B^{\alpha(k+m-1)\dot\alpha(k-m+1)} \nonumber
\\
 && + (k-m)\beta^{+-}_{k,m} e_\alpha{}^{\dot\alpha}
B^{\alpha(k+m+1)\dot\alpha(k-m-1)}.
\end{eqnarray} 
Then one has to determine the coefficients $\alpha^{ij}_{k,m}$,
$\beta^{ij}_{k,m}$. Such problem arise three times for bosons and,
similarly, three times for fermions. Namely, the calculation of the
right coefficients in expressions for 2-curvatures, the derivation of
the linear relations for 2-curvatures and the derivation of the
unfolded equations can be reduced to the problem stated above, with
additional restrictions on the coefficients (for example, the
normalization choice $\alpha^{+-}_{k,m}=1$ or the hermiticity
condition $\alpha^{-+}_{k,m}=\alpha^{+-}_{k,m}$). It is thus important
to solve this problem once in the general case. It immediately follows
that $\alpha^{ij}_{k,m}=\beta^{ij}_{k,m}$. Then, the following
recurrent relations for $\alpha^{ij}_{k,m}$ hold:
\begin{eqnarray}
(k-m)\big[\alpha^{--}_{k,m}\alpha^{++}_{k-1,m} +
\alpha^{-+}_{k,m+1}\alpha^{+-}_{k,m}\big] + 2\lambda^2 &=&
(k-m+2)\big[
\alpha^{--}_{k+1,m}\alpha^{++}_{k,m} + \alpha^{+-}_{k,m-1}
\alpha^{-+}_{k,m}\big], \nonumber
\\
(k+m)\big[\alpha^{--}_{k,m}\alpha^{++}_{k-1,m} +
\alpha^{-+}_{k,m}\alpha^{+-}_{k,m-1}\big] + 2\lambda^2 &=&
(k+m+2)\big[ \alpha^{--}_{k+1,m}\alpha^{++}_{k,m} + 
\alpha^{-+}_{k,m+1} \alpha^{+-}_{k,m}\big], \nonumber
\\
(k+m+2)\alpha^{-+}_{k+1,m}\alpha^{++}_{k,m} &=&
(k+m)\alpha^{-+}_{k,m}\alpha^{++}_{k,m-1}, \nonumber
\\
(k-m+2)\alpha^{+-}_{k+1,m}\alpha^{++}_{k,m} &=&
(k-m)\alpha^{+-}_{k,m}\alpha^{++}_{k,m+1}, 
\\
(k-m)\alpha^{--}_{k,m}\alpha^{-+}_{k-1,m} &=& (k-m+2)
\alpha^{--}_{k,m-1}\alpha^{-+}_{k,m}, \nonumber
\\
(k+m)\alpha^{--}_{k,m}\alpha^{+-}_{k-1,m} &=& (k+m+2)
\alpha^{--}_{k,m+1}\alpha^{+-}_{k,m}. \nonumber
\end{eqnarray}
The coefficients $\alpha^{ij}_{k,m}$ satisfy those relations iff:
\begin{eqnarray}
\alpha^{+-}_{k,m}\alpha^{-+}_{k,m+1} &=& 
\frac{A_m}{(k-m)(k-m+1)(k+m+1)(k+m+2)}, \nonumber
\\
\alpha^{--}_{k+1,m}\alpha^{++}_{k,m} &=&
\frac{A_{k+1}}{(k-m+1)(k-m+2)(k+m+1)(k+m+2)},
\\
A_{m} &=& C_1+C_2(m+1)m+(m+1)^2m^2\lambda^2. \nonumber
\end{eqnarray}
Note that the fermionic coefficients $\alpha^{ij}_{k,m}$ are redefined
as $\alpha^{ij}_{k-1/2,m-1/2}$ in  the main text. The normalization of
the coefficients $\alpha^{ij}_{k,m}$ and the constants $C_1,C_2$ are
determined from the additional restrictions. Also note that the given
expressions are not applicable in case of $m=\pm k$. In particular,
this explains why the expressions for $ R^{\alpha(2k)}$ significantly
differ from the general case $ R^{\alpha(k+m){\dot{\alpha}}(k-m)}$.

Now let us discuss the relations between curvatures. In case of the
free field, each curvature is linear on the fields and has the form:
\begin{equation}
R^A = D W^A + F^A(W^B), \qquad
F^A(W^B) = \sum_{B\in B(A)} f^A{}_B W^{B},
\end{equation}
The exterior derivative of the curvature $R^A$ hence can be expressed
in terms of curvatures which contain the derivatives of the fields
$W^B$, $B\in B(A)$:
\begin{equation}
DR^A = \sum_{B\in B(A)} (-1)^{\deg f^A{}_B}f^A{}_B R(B) + G^A(W^B).
\end{equation}
Here $G^A(W^B)$ does not contain the exterior derivatives of the
fields. The factor $ (-1)^{\deg f^A{}_B}$ is due to anticommutativity
of the exterior product and the exterior derivative. The gauge
invariance implies $G^A(W^B)\equiv 0$, which means that the derivative
of the curvature is expressed via other curvatures. Indeed, one can
check by straightforward calculation that:
\begin{eqnarray}
0 &=& D R^{\alpha(k+m)\dot\alpha(k-m)}
+ (k+m)(k-m)\alpha^{++}_{k,m} e^{\alpha\dot\alpha}
R^{\alpha(k+m-1)\dot\alpha(k-m-1)} \nonumber
\\
 && + \alpha^{--}_{k,m} e_{\alpha\dot\alpha}
R^{\alpha(k+m+1)\dot\alpha(k-m+1)}
+ (k+m)\alpha^{-+}_{k,m} e^\alpha{}_{\dot\alpha}
R^{\alpha(k+m-1)\dot\alpha(k-m+1)} \nonumber
\\
 && + (k-m)\alpha^{+-}_{k,m} e_\alpha{}^{\dot\alpha}
R^{\alpha(k+m+1)\dot\alpha(k-m-1)}, \nonumber
\\
0 &=& D R^{\alpha(2k)}
+ e_{\alpha\dot\alpha} R^{\alpha(2k+1)\dot\alpha}
+ 2k\alpha^{-+}_{k,k} e^\alpha{}_{\dot\alpha}
R^{\alpha(2k-1)\dot\alpha} \nonumber
\\
 && + 4k(2k-1)\alpha^{-+}_{k,k}\alpha^{--}_{k,k-1} E^{\alpha(2)}
C^{\alpha(2k-2)} + \alpha^{+-}_{k+1,k} E_{\alpha(2)} C^{\alpha(2k+2)},
\\
0 &=& D C^{\alpha(k+m)\dot\alpha(k-m)} + 
R^{\alpha(k+m)\dot\alpha(k-m)} \nonumber
\\
 && + (k+m)(k-m)\alpha^{--}_{k,m} e^{\alpha\dot\alpha}
C^{\alpha(k+m-1)\dot\alpha(k-m-1)}
+ \alpha^{++}_{k,m} e_{\alpha\dot\alpha}
C^{\alpha(k+m+1)\dot\alpha(k-m+1)} \nonumber
\\
 && + (k+m) e^\alpha{}_{\dot\alpha}
C^{\alpha(k+m-1)\dot\alpha(k-m+1)}
+ (k-m)\alpha^{-+}_{k,m} e_\alpha{}^{\dot\alpha}
C^{\alpha(k+m+1)\dot\alpha(k-m-1)}. \nonumber
\end{eqnarray}
Here, in the last equality one has to omit the terms with the factor
$(k\mp m)$ in case of $m=\pm k$. For the lowest bosonic curvatures,
the expressions are slightly different:
\begin{eqnarray}
0 &=& D R^{\alpha(2)} + 2\beta_2 e^\alpha{}_{\dot\alpha}
R^{\alpha\dot\alpha} + 3\mu_2 e_{\alpha\dot\alpha} 
R^{\alpha(3)\dot\alpha} + \mu_{1}{}^{2}E^\alpha{}_\beta 
C^{\alpha\beta} + 2\beta_2 \mu_1 E^{\alpha(2)} C
- 6\mu_2 E_{\alpha(2)} C^{\alpha(4)}, \nonumber
\\
0 &=& D C^{\alpha\dot\alpha} + R^{\alpha\dot\alpha} +
e^\alpha{}_{\dot\beta} C^{\dot\alpha\dot\beta} +
e_\beta{}^{\dot\alpha} C^{\alpha\beta} + \frac{\mu_1}{2}
e^{\alpha\dot\alpha} C + \mu_2 e_{\alpha\dot\alpha}
C^{\alpha(2)\dot\alpha(2)}, \nonumber
\\
0 &=& D C^{\alpha(2)} + R^{\alpha(2)} + 2\beta_2
e^\alpha{}_{\dot\alpha} C^{\alpha\dot\alpha}
+ 3\mu_2 e_{\alpha\dot\alpha} C^{\alpha(3)\dot\alpha},
\\
0 &=& DR + \mu_1 e_{\alpha\dot\alpha} R^{\alpha\dot\alpha} + 2\mu_1
E_{\alpha(2)} C^{\alpha(2)} + 2\mu_1 E_{\dot\alpha(2)}
R^{\dot\alpha(2)}, \nonumber
\\
0 &=& DC + R + \mu_1 e_{\alpha\dot\alpha} C^{\alpha\dot\alpha}.
\nonumber
\end{eqnarray}
It is possible therefore to obtain the following relations for the
derivative of the product of the two curvatures:
\begin{eqnarray}
\label{eqties}
&& - D( R^{\alpha(k+m)\dot\alpha(k-m)}
C_{\alpha(k+m)\dot\alpha(k-m)})
= R^{\alpha(k+m)\dot\alpha(k-m)}
R_{\alpha(k+m)\dot\alpha(k-m)} \nonumber
\\
 && + (k+m)(k-m)\alpha^{--}_{k,m}
\big[ R^{\alpha(k+m-1)\dot\alpha(k-m-1)} e^{\alpha\dot\alpha}
C_{\alpha(k+m)\dot\alpha(k-m)} \nonumber
\\
 && \qquad \qquad \qquad \qquad
+ R^{\alpha(k+m)\dot\alpha(k-m)} e_{\alpha\dot\alpha}
C_{\alpha(k+m-1)\dot\alpha(k-m+1)} \big] \nonumber
\\
&& + (k+m)\alpha^{-+}_{k,m} \big[ R^{\alpha(k+m-1)\dot\alpha(k-m+1)}
e^\alpha{}_{\dot\alpha} C_{\alpha(k+m)\dot\alpha(k-m)} \nonumber
\\
 && \qquad \qquad \qquad
- R^{\alpha(k+m)\dot\alpha(k-m)} e_\alpha{}^{\dot\alpha}
C_{\alpha(k+m-1)\dot\alpha(k-m+1)} \big] \nonumber
\\
 && + (k-m) \big[ R^{\alpha(k+m+1)\dot\alpha(k-m-1)}
e_\alpha{}^{\dot\alpha} C_{\alpha(k+m)\dot\alpha(k-m)} \nonumber
\\
 && \qquad \qquad 
- R^{\alpha(k+m)\dot\alpha(k-m)} e^\alpha{}_{\dot\alpha}
C_{\alpha(k+m+1)\dot\alpha(k-m-1)} \big] \nonumber
\\
 && + \alpha^{++}_{k,m} \big[ R^{\alpha(k+m+1)\dot\alpha(k-m+1)}
e_{\alpha\dot\alpha} C_{\alpha(k+m)\dot\alpha(k-m)} \nonumber
\\
 && \qquad 
+ R^{\alpha(k+m)\dot\alpha(k-m)} e^{\alpha\dot\alpha}
C_{\alpha(k+m+1)\dot\alpha(k-m-1)} \big], 
\end{eqnarray}
\begin{eqnarray}
&& - D( R^{\alpha(2k)} C_{\alpha(2k)} ) = \alpha_{k,k}{}^{++}
R^{\alpha(2k+1)\dot\alpha} e_{\alpha\dot\alpha} C_{\alpha(2k)}
+ 2k\alpha^{-+}_{k,k} R^{\alpha(2k-1)\dot\alpha}
e^\alpha{}_{\dot\alpha} C_{\alpha(2k)} \nonumber
\\
 && + 2\alpha^{++}_{k,k} C^{\alpha(2k+2)} E_{\alpha(2)} C_{\alpha(2k)}
+ 4k(2k-1)\alpha^{-+}_{k,k}\alpha^{--}_{k,k-1} C^{\alpha(2k-2)}
E^{\alpha(2)} C_{\alpha(2k)} \nonumber
\\
 && + \frac{\alpha^{++}_{k-1}\alpha^{--}_{k}}{k+1} 
C^{\alpha(2k-1)\beta} E^\alpha{}_\beta C_{\alpha(2k)}
+ R^{\alpha(2k)} R_{\alpha(2k)}
+ \alpha_{k,k}^{++} R^{\alpha(2k)} e^{\alpha\dot\alpha}
C_{\alpha(2k+1)\dot\alpha} \nonumber
\\
 && - 2k\alpha^{-+}_{k,k} R^{\alpha(2k)} e_\alpha{}^{\dot\alpha}
C_{\alpha(2k-1)\dot\alpha}, 
\end{eqnarray}
\begin{eqnarray}
&& - D( C^{\alpha(k+m)\dot\alpha(k-m)} e^\alpha{}_{\dot\alpha}
C_{\alpha(k+m+1)\dot\alpha(k-m-1)}) = \nonumber
\\
 && R^{\alpha(k+m)\dot\alpha(k-m)} e^\alpha{}_{\dot\alpha} 
C_{\alpha(k+m+1)\dot\alpha(k-m-1)} 
+ R^{\alpha(k+m+1)\dot\alpha(k-m-1)} e_\alpha{}^{\dot\alpha}
C_{\alpha(k+m)\dot\alpha(k-m)} \nonumber
\\
 && + (k+m)(k-m+1)\alpha^{--}_{k,m} C^{\alpha(k+m-1)\dot\alpha(k-m-1)}
E^{\alpha(2)} C_{\alpha(k+m+1)\dot\alpha(k-m-1)} \nonumber
\\
 && - (k+m+2)(k-m-1)\alpha^{--}_{k,m+1} C^{\alpha(k+m)\dot\alpha(k-m)}
E_{\dot\alpha(2)} C_{\alpha(k+m)\dot\alpha(k-m-2)} \nonumber
\\
 && - \alpha^{++}_{k,m} C^{\alpha(k+m+1)\dot\alpha(k-m+1)}
E_{\dot\alpha(2)} C_{\alpha(k+m+1)\dot\alpha(k-m-1)} \nonumber
\\
 && + \alpha^{++}_{k,m+1} C^{\alpha(k+m)\dot\alpha(k-m)} E^{\alpha(2)}
C_{\alpha(k+m+2)\dot\alpha(k-m)} \nonumber
\\
 && - (k-m-1) C^{\alpha(k+m+1)\dot\alpha(k-m-2)\dot\beta}
E_{\dot\beta}{}^{\dot\alpha} C_{\alpha(k+m+1)\dot\alpha(k-m-1)}
\nonumber
\\
 && + (k-m+1) C^{\alpha(k+m)\beta\dot\alpha(k-m-1)}
E^\alpha{}_\beta C_{\alpha(k+m+1)\dot\alpha(k-m-1)} \nonumber
\\
 && - (k+m+2)\alpha^{-+}_{k,m+1} C^{\alpha(k+m)\dot\alpha(k-m)}
E_{\dot\alpha}{}^{\dot\beta}
C_{\alpha(k+m)\dot\alpha(k-m-1)\dot\beta} \nonumber
\\
 && + (k+m)\alpha^{-+}_{k,m+1} C^{\alpha(k+m)\dot\alpha(k-m)}
E^\beta{}_\alpha C_{\alpha(k+m-1)\beta\dot\alpha(k-m)}, 
\end{eqnarray}
\begin{eqnarray}
&& - D( C^{\alpha(k+m)\dot\alpha(k-m)} e^{\alpha\dot\alpha}
C_{\alpha(k+m+1)\dot\alpha(k-m+1)}) = \nonumber
\\
 &&  R^{\alpha(k+m)\dot\alpha(k-m)} e^{\alpha\dot\alpha} 
C_{\alpha(k+m+1)\dot\alpha(k-m+1)}
- R^{\alpha(k+m+1)\dot\alpha(k-m+1)} e_{\alpha\dot\alpha}
C_{\alpha(k+m)\dot\alpha(k-m)} \nonumber
\\
 && - \alpha^{++}_{k,m}\big[ C^{\alpha(k+m)\beta\dot\alpha(k-m+1)}
E^\alpha{}_\beta C_{\alpha(k+m+1)\dot\alpha(k-m+1)} \nonumber
\\
 && + C^{\alpha(k+m+1)\dot\beta\dot\alpha(k-m)}
E^{\dot\alpha}{}_{\dot\beta} C_{\alpha(k+m+1)\dot\alpha(k-m+1)} \big]
\nonumber
\\
 && - (k+m)\alpha^{-+}_{k,m} C^{\alpha(k+m-1)\dot\alpha(k-m+1)}
E^{\alpha(2)} C_{\alpha(k+m+1)\dot\alpha(k-m+1)} \nonumber
\\
 && + (k+m+2)\alpha^{-+}_{k+1,m} C^{\alpha(k+m)\dot\alpha(k-m)}
E^{\dot\alpha(2)} C_{\alpha(k+m)\dot\alpha(k-m+2)} \nonumber
\\
 && - (k-m) C^{\alpha(k+m+1)\dot\alpha(k-m-1)} E^{\dot\alpha(2)}
C_{\alpha(k+m+1)\dot\alpha(k-m+1)} \nonumber
\\
 && + (k-m+2) C^{\alpha(k+m)\dot\alpha(k-m)} E^{\alpha(2)}
C_{\alpha(k+m+2)\dot\alpha(k-m)} \nonumber
\\
 && - (k+m)(k-m+2)\alpha^{--}_{k+1,m} C^{\alpha(k+m)\dot\alpha(k-m)}
E^\beta{}_\alpha C_{\alpha(k+m-1)\beta\dot\alpha(k-m)} \nonumber
\\
 && - (k+m+2)(k-m)\alpha^{--}_{k+1,m} C^{\alpha(k+m)\dot\alpha(k-m)}
E^{\dot\beta}{}_{\dot\alpha} 
C_{\alpha(k+m)\dot\beta\dot\alpha(k-m+1)}.
\label{eqties1}
\end{eqnarray}

Those relations determine the arbitrariness of the coefficients in the
Lagrangian. Namely, the explicit expressions for their shifts are:
\begin{eqnarray}
\label{genbs}
\pm a_{k,\pm m} &=& \pm a^{(0)}_{k,\pm m}+p_{k,m}, \nonumber
\\
\pm b_{k,\pm m} &=& \pm b^{(0)}_{k,\pm
m}+p_{k,m}\alpha_{k,m}^{++}-p_{k+1,m}(k+m+1)(k-m+1)\alpha^{--}_{k+1,m}+r_{k,m}, \nonumber
\\
\mp c_{k+1,\pm m} &=& \mp c^{(0)}_{k+1,\pm m}+p_{k,m}
\alpha_{k,m}^{++}-p_{k+1,m}(k+m+1)(k-m+1)\alpha^{--}_{k+1,m}+r_{k,m},
\nonumber
\\
d_{k,m} &=& d^{(0)}_{k,m}
-(k-m)p_{k,m}+(k+m+1)\alpha^{-+}_{k,m+1}p_{k,m+1}+q_{k,m}, \nonumber
\\
d_{k,-1-m} &=& d^{(0)}_{k,-1-m}-(k-m)p_{k,m}
+(k+m+1)\alpha^{-+}_{k,m+1}p_{k,m+1}-q_{k,m}, \nonumber
\\
e_{k,m} &=& e^{(0)}_{k,m} +
(k+m)\alpha^{-+}_{k,m+1}q_{k,m}+(k-m+2)q_{k,m-1}, \nonumber
\\
 && -
(k-m+2)(k+m)\alpha^{--}_{k+1,m}r_{k,m}+\alpha^{++}_{k-1,m}r_{k-1,m},
\nonumber
\\
e_{k,0} &=& e^{(0)}_{k,0} + k\alpha^{-+}_{k,1}q_{k,0} -
k(k+2)\alpha^{--}_{k+1,0}r_{k,0}+\alpha^{++}_{k-1,0}r_{k-1,0},
\nonumber
\\
e_{k,-m} &=& e^{(0)}_{k,-m} +
(k+m+2)\alpha^{-+}_{k,m+1}q_{k,m}+(k-m)q_{k,m-1}, \nonumber
\\
 && +
(k+m+2)(k-m)\alpha^{--}_{k+1,m}r_{k,m}-\alpha^{++}_{k-1,m}r_{k-1,m},
\nonumber
\\
e_{1,0} &=& e^{(0)}_{1,0} +4\beta_2q_{1,0}-3\alpha^{--}_{2,0}r_{1,0},
\\
e_{k,k} &=&
e^{(0)}_{k,k}+\frac{\alpha^{++}_{k-1}\alpha^{--}_{k}}{k+1}p_{k,k}+2q_{k,k-1}-4k\alpha^{--}_{k+1,k}r_{k,k}, \nonumber
\\
e_{1,1} &=& e^{(0)}_{1,1}+\mu_1^2
p_{1,1}+2q_{1,0}-4\alpha^{--}_{2,1}r_{1,1}, \nonumber
\\
f_{k,m} &=&
f^{(0)}_{k,m}-(k+m+1)(k-m+2)\alpha^{--}_{k+1,m}q_{k+1,m}+
\alpha_{k,m+1}^{++}q_{k,m}, \nonumber
\\
 && -(k+m+1)\alpha^{-+}_{k,m+1}r_{k,m+1}+(k-m+2)r_{k,m}, \nonumber
\\
f_{k,-m} &=& f^{(0)}_{k,-m}+(k+m+2)(k-m+1)\alpha^{--}_{k+1,m}
q_{k+1,m}-\alpha_{k,m-1}^{++}q_{k,m-1}, \nonumber
\\
 && - (k+m+2)\alpha^{-+}_{k+1,m}r_{k,m}+(k-m+1)r_{k,m-1}, \nonumber
\\
f_{k,k} &=&
f^{(0)}_{k,k}-2\alpha^{++}_{k,k}p_{k,k}-4(k+1)(2k+1)\alpha^{-+}_{k+1,k+1}\alpha^{--}_{k+1,k}p_{k+1,k+1}, \nonumber
\\
 && - 2(2k+1)\alpha^{--}_{k+1,k}q_{k+1,k}+2r_{k,k}, \nonumber
\\
f_{0,0} &=& f^{(0)}_{0,0}-\mu_1q_{1,0}-2\beta_2
\mu_1p_{1,1}+2\mu_1p_{0,0}+2r_{0,0}. \nonumber
\end{eqnarray}

The values of the non-zero shift parameters $p_{k,m},q_{k,m}$ (all
$r_{k,m}=0$) which gives the solution for the partially massless
bosonic cases:
\begin{eqnarray}
p_{k,m} &=& - \frac{(k-1)!(k+m+1)!k!}{(k-m)!^2(k-m+1)!
\prod_{i=1}^{m}\alpha^{-+}_{i}}, \quad m \ge n, \nonumber
\\
q_{k,n-1} &=& - \frac{(k-1)!(k+n)!k!}{(k-n)!(k-n+1)!(k-n+2)!
\prod_{i=1}^{n-1}\alpha^{-+}_{i}}.
\end{eqnarray}

The explicit expressions for the shifts of the Lagrangian parameters
for the fermionic case:
\begin{eqnarray}
\label{genfs}
a_{k,\pm m} &=& a^{(0)}_{k,m}+p_{k,m}, \nonumber
\\
b_{k,\pm m} &=& b^{(0)}_{k,m}+p_{k,m}\alpha_{k,m}^{++}-
p_{k+1,m}(k+m+2)(k-m+1)\alpha^{--}_{k+1,m}+r_{k,m}, \nonumber
\\
c_{k+1,\pm m} &=& c^{(0)}_{k,m}-p_{k,m}\alpha_{k,m}^{++}+p_{k+1,m}
(k+m+2)(k-m+1)\alpha^{--}_{k+1,m}-r_{k,m}, \nonumber
\\
d_{k,m} &=& d^{(0)}_{k,m}
-(k-m)p_{k,m}+(k+m+2)\alpha^{-+}_{k,m+1}p_{k,m+1}+q_{k,m}, \nonumber
\\
d_{k,-1} &=& d^{(0)}_{k,-1}+2q_{k,-1}, \nonumber
\\
d_{k,-m} &=&
d^{(0)}_{k,-m}+(k-m+2)p_{k,m-2}-(k+m)\alpha^{-+}_{k,m-1}p_{k,m-1}+q_{k,m-2}, \nonumber
\\
e_{k,m} &=& e^{(0)}_{k,m} +
(k+m+1)\alpha^{-+}_{k,m+1}q_{k,m}+(k-m+2)q_{k,m-1}, \nonumber
\\
 && -
(k-m+2)(k+m+1)\alpha^{--}_{k+1,m}r_{k,m}+\alpha^{++}_{k-1,m}r_{k-1,m},
\nonumber
\\
e_{k,-m} &=& e^{(0)}_{k,-m} -
(k+m+2)\alpha^{-+}_{k,m}q_{k,m-1}-(k-m+1)q_{k,m-2},
\\
 && +
(k-m+3)(k+m)\alpha^{--}_{k+1,m-1}r_{k,m-1}-\alpha^{++}_{k-1,m-1}r_{k-1,m-1}, \nonumber
\\
e_{k,k} &=& e^{(0)}_{k,k}+\frac{2\alpha^{++}_{k-1}\alpha^{--}_{k}}
{2k+3}p_{k,k}+2q_{k,k-1}-2(2k+1)\alpha^{--}_{k+1,k}r_{k,k}, \nonumber
\\
f_{k,m} &=&
f^{(0)}_{k,m}-(k+m+2)(k-m+2)\alpha^{--}_{k+1,m}q_{k+1,m}+
\alpha_{k,m+1}^{++}q_{k,m}, \nonumber
\\
 && - (k+m+1)\alpha^{-+}_{k,m+1}r_{k,m+1}+(k-m+2)r_{k,m}, \nonumber
\\
f_{k,-m} &=& f^{(0)}_{k,-m}+(k+m+1)(k-m+3)\alpha^{--}_{k+1,m-1}
q_{k+1,m-1}-\alpha_{k,m}^{++}q_{k,m-1}, \nonumber
\\
 && + (k+m)\alpha^{-+}_{k,m}r_{k,m}-(k-m+1)r_{k,m-1}, \nonumber
\\
f_{k,k} &=&
f^{(0)}_{k,k}-2\alpha^{++}_{k,k}p_{k,k}-4(k+1)(2k+3)\alpha^{-+}_{k+1,k+1}\alpha^{--}_{k+1,k}p_{k+1,k+1}, \nonumber
\\
 && - 4(k+1)\alpha^{--}_{k+1,k}q_{k+1,k}+2r_{k,k}. \nonumber
\end{eqnarray}

The values of the non-zero shift parameters $p_{k,m},q_{k,m}$ (all
$r_{k,m}=0$) which gives the solution for the partially massless
fermionic cases:
\begin{eqnarray}
p_{k,m} &=& -
\frac{(k+m+2)!k!^2}{4(k-m)!^2(k-m+1)!\prod_{i=0}^{m}\alpha^{-+}_{i}},
\qquad m\ge n, \nonumber
\\
q_{k,n-1} &=& - \frac{(k+n+1)!k!^2}{4(k-n)!(k-n+1)!(k-n+2)!
\prod_{i=0}^{n-1}\alpha^{-+}_{i}}.
\end{eqnarray}

\end{document}